\documentclass{aa}
 \usepackage[varg]{txfonts}
\usepackage[english]{babel}
\usepackage{booktabs}
\usepackage{graphicx}
\usepackage{caption}
\usepackage{fixltx2e}
\usepackage{amsmath,amssymb}
\usepackage{mathabx}
\usepackage{url}
\usepackage{rotating}
\usepackage{natbib}
\usepackage{bm}
\usepackage{varioref}
\usepackage{longtable}
\usepackage{supertabular,booktabs}
\usepackage{xcolor}

\title{The dynamics of three nearby E0 galaxies in refracted gravity}

 \author{V.~Cesare\inst{1,2,3}, A.~Diaferio\inst{1,2}, and T.~Matsakos\inst{1}}
 \institute{Dipartimento di Fisica, Università di Torino, via P. Giuria 1, 10125 Torino, Italy, \email{valentina.cesare@unito.it}
 	\and Istituto Nazionale di Fisica Nucleare (INFN), Sezione di Torino, Torino, Italy 
 	\and Istituto Nazionale di Astrofisica (INAF), Osservatorio Astrofisico di Catania, via Santa Sofia 78, 95123 Catania, Italy, \email{valentina.cesare@inaf.it}}
 \date{Received 24 February 2021 / Accepted 12 October 2021}

\abstract
{We tested whether refracted gravity, a theory of modified gravity that describes the dynamics of galaxies without the aid of dark matter, can model the dynamics of the three massive elliptical galaxies, NGC 1407, NGC 4486, and NGC 5846, out to $\sim$$10R_{\rm e}$, where the baryonic mass component fades out and dark matter is required in Newtonian gravity. We probed these outer regions with the kinematics of the globular clusters provided by the SLUGGS survey. Refracted gravity mimics dark matter with the gravitational permittivity, a monotonic function of the local mass density depending on three parameters, $\epsilon_0$, $\rho_{\rm c}$, and $Q$, which are expected to be universal. Refracted gravity satisfactorily reproduces the velocity dispersion profiles of the stars and red and blue globular clusters, with stellar mass-to-light ratios in agreement with stellar population synthesis models, and orbital anisotropy parameters consistent with previous results obtained in Newtonian gravity with dark matter. The sets of the three parameters of the gravitational permittivity found for each galaxy are consistent with each other within $\sim$2$\sigma$. We compare the mean $\{\epsilon_0,Q,\log_{10}\left[\rho_{\rm c}\left(\mathrm{g}\, \mathrm{cm}^{-3}\right)\right]\} = \{0.089^{+0.038}_{-0.035}, 0.47^{+0.29}_{-0.21}, -24.25^{+0.28}_{-0.20}\}$ found here with the means of the parameters required to model the rotation curves and vertical velocity dispersion profiles of 30 disk galaxies from the DiskMass Survey (DMS): $\rho_{\rm c}$ and $Q$ agree within 1$\sigma$ with the DMS values, whereas $\epsilon_0$ agrees within 3$\sigma$. 
This agreement suggests that ellipticals and disk galaxies allow for common values of the parameters of the permittivity and supports the universality of the permittivity function. 
}


\begin{document}
	\maketitle
	\titlerunning
	\authorrunning
	
   \section{Introduction}
   \label{sec:Introduction}

   The observed kinematical properties of the baryonic matter in galaxies suggest that galaxies are embedded within extended halos of dark matter whose distribution and contribution to the total mass of the galaxy largely vary with the size and the morphological type of the galaxy~\citep{DelPopolo_2014}. This result consistently fits within the standard cosmological model, where the properties of the Universe on large scales, including the temperature anisotropies of the cosmic microwave background (CMB)~\citep{Planck_2018_V}, the large-scale distribution of galaxies~\citep{Dodelson_2002}, and the abundance of light elements~\citep{Cyburt_2016}, imply that the matter content of the Universe is dominated by cold dark matter. 

  The standard cosmological model, however, suffers from some tensions both on large scales and on the scale of galaxies. On large scales, the discrepancy between the value of the Hubble constant, $H_0$, derived from observations in the local Universe and the value derived from the CMB~\citep{Verde_2019}, the unlikely features in the CMB temperature anisotropies~\citep{Schwarz_2016}, and the deficiency of $^7$Li~\citep{Mathews_2019} have recently challenged the standard model. On the scale of galaxies, the tensions have been long-lasting: in disk galaxies, fine-tuned interactions between baryonic and dark matter are necessary to describe the radial acceleration relation (RAR) or the baryonic Tully-Fisher relation~\citep{McGaugh_2020}; the missing satellite problem, the plane of satellite galaxies around the Milky Way, and the cusp-core problem in dwarf galaxies pose additional challenges~\citep[see e.g.][for reviews]{Kroupa_2012,DelPopolo_and_LeDelliou_2017,deMartino_2020}. Moreover, the search for the elementary particles constituting dark matter remains inconclusive~\citep{Tanabashi_2018}. 

  On the scale of galaxies, a number of models beyond the standard model have been suggested. Dark matter particles might not be cold and collisionless, but rather self-interacting~\citep{Tulin_and_Yu_2018}, fuzzy~\citep{Hui_2017}, axion-like~\citep{Marsh_2016}, or a superfluid~\citep{Ferreira_2019}. Alternatively, the theory of gravity might break down on galaxy and larger cosmic scales and thus dark matter might be unnecessary. Modified gravity theories proposed in the literature are numerous: for example, $f(R)$ gravity, where the Ricci scalar $R$ in the Einstein-Hilbert action is replaced by an arbitrary function of $R$~\citep{Nojiri_and_Odintsov_2011} or conformal gravity, where $R$ is replaced by the contraction of the fourth-rank Weyl tensor~\citep{Mannheim_2019}. However, both modified dark matter models and modified gravity theories have encountered a number of challenges. For example, neither $f(R)$ nor conformal gravity appear to be able to yield the correct abundance of light elements~\citep{Azevedo_and_Avelino_2018,Elizondo_and_Yepes_1994}. Conformal gravity also appears to suffer from  severe fine-tuning in the description of the kinematics of disk galaxies~\citep{Campigotto_2019} and requires an intergalactic medium in galaxy clusters that is too hot~\citep{Horne_2006,Diaferio_and_Ostorero_2009}. MOdified Newtonian Dynamics (MOND) has been proved to be the most successful phenomenological model on the scale of galaxies~\citep{Sanders_and_McGaugh_2002, McGaugh_2020}: attempts to build a covariant formulation of MOND have been challenging~\citep{Famaey_and_McGaugh_2012,Milgrom_2015, Hernandez_2019}, but recent developments appear to provide directions for viable solutions~\citep{Skordis_and_Zlosnik_2019,Skordis_and_Zlosnik_2020}.

  \citet{Matsakos_and_Diaferio_2016} suggested refracted gravity (RG), a theory of gravity where the Poisson equation is modified by the introduction of the gravitational permittivity, a monotonic increasing function of the local mass density. Refracted gravity appears to properly describe the observed phenomenology on the scale of galaxies and clusters of galaxies, with the role of dark matter mimicked by the gravitational permittivity~\citep{Matsakos_and_Diaferio_2016,Cesare_2020}. In addition, RG provides a straightfoward covariant extension within the family of the scalar-tensor theories~\citep{Sanna_2021}. In the covariant extension, the role of the gravitational permittivity is played by the scalar field. This same scalar field is also responsible for the accelerated expansion of the Universe. Thus, RG has the attractive feature of unifying the two dark sectors of the Universe into a single scalar field~\citep{Carneiro_2018,Bertacca_2010}. In addition, in its ultra-weak field limit, the covariant extension of RG seems to suggest the emergence of an acceleration scale below which Newtonian gravity breaks down. This acceleration scale, originally proposed by \citet{Milgrom_1983a} in MOND, is strongly supported by the observed dynamics on the scale of galaxies~\citep[see e.g.][for reviews]{deMartino_2020,Merritt_2020}. However, the emergence of an acceleration scale in the covariant extension of RG still requires a solid confirmation.

  In RG, the presence of regions where the local mass density is sufficiently low distinctively affects the gravitational field: in flat systems, such as disk galaxies, the gravitational permittivity bends the lines of the gravitational field towards the system, thus determining a boost of the gravitational field in the plane perpendicular to the minor axis of the system. In spherically symmetric systems, the field lines remain radial, but the gravitational field is enhanced by the inverse of the gravitational permittivity. In \citet{Cesare_2020}, we investigated the dynamics of 30 disk galaxies from the DiskMass Survey (DMS,~\citealt{DMSI_2010a}). These galaxies are almost face-on, with an inclination angle with respect to the line-of-sight in the range $6-46$ degrees. This inclination enables the measurements of both the rotation curve and the profile of the stellar velocity dispersion perpendicular to the plane of the disk, although the measures of the rotation velocities are generally less accurate than for edge-on galaxies. \citet{Cesare_2020} proved that the bending of the gravitational field lines occurring in RG is sufficient to describe the kinematics of these 30 DMS galaxies without resorting to dark matter. This description depends on three parameters appearing in the gravitational permittivity. A single set of parameters  describes the entire sample of 30 galaxies, supporting the expectation that these parameters should be universal. These encouraging results on flat systems motivated us to test whether RG is also able to describe the dynamics of spherical systems with the same set of parameters. In other words, we wish to probe whether the boost of the gravitational field can be set by the gravitational permittivity alone, independently of the redirection of the field lines.  
  
  Here, we illustrate the result of this test on three E0 galaxies, which are approximately spherical. We considered \object{NGC 1407}, \object{NGC 4486}, alias \object{M87}, and \object{NGC 5846} from the SLUGGS survey,\footnote{\url{https://sluggs.swin.edu.au/Start.html.}}  a spectrophotometric survey of more than 4000 extragalactic globular clusters (GCs) around 27 early-type galaxies \citep{Pota_2013,Brodie_2014,Forbes_2017}. To our knowledge, SLUGGS is the only survey with publicly available photometric and kinematic information at large radii of early-type galaxies. The three galaxies we consider here are the only E0 galaxies in the SLUGGS survey. In elliptical galaxies, the baryonic matter within $R_{\rm e}$, the effective radius at half projected luminosity, suffices to describe the galaxy dynamics with Newtonian gravity~\citep[e.g.][]{Dekel_2005,Mamon_and_Lokas_2005,Proctor_2009,ATLAS_XV}. Thus, to probe RG, we need kinematic information in the outer regions, where Newtonian gravity breaks down unless dark matter is assumed to exist. X-ray emitting gas, planetary nebulae, and globular clusters (GCs) have been adopted as kinematic tracers in these regions, out to $\sim$$10R_{\rm e}$, where the stellar luminosity fades out~\citep{Danziger_1997,Mathews_and_Brighenti_2003,Werner_2019,Pota_2013, ePNS_I_2018}. In addition, GCs usually appear separated into two distinct populations, according to their colour. The red sub-population of the GC system is generally more spatially concentrated and has a smaller velocity dispersion than the blue sub-population~\citep{Geisler_1996,Ashman_and_Zepf_1998,Bassino_2006,Brodie_and_Strader_2006,Faifer_2011,Strader_2011,Forbes_2012,Lee_2008,Pota_2013}. By adopting the SLUGGS survey and thus the GCs as tracers of the velocity field in the outer regions of ellipticals, we actually have two tracers rather than one and thus we can more strongly constrain the model. Therefore, the three E0 galaxies of our sample provide a suitable dataset for our test of RG.
  
  Studying the outer regions of ellipticals might be, however, particularly insidious. Ellipticals tend to live in dense environments: NGC 1407 and NGC 5846 are within groups and NGC 4486 is the central galaxy of the Virgo cluster. Therefore, their outer regions are subject to the gravitational field of nearby galaxies and of the hosting system as a whole. Nevertheless, in our model below we neglect the effects of the environment and we model each galaxy as an isolated system. Despite this severe assumption, our RG model can still describe reasonably well the overall dynamics of these elliptical galaxies.

  Section~\ref{sec:RG} summarises the main features of RG. Sections~\ref{sec:Photometric_data} and~\ref{sec:Spectroscopic_data} describe the photometric and spectroscopic data that we used in our analysis. In Sect.~\ref{sec:Mass_modelling}, we illustrate our RG dynamical model, and in Sect.~\ref{sec:Results} we report our results. We conclude in Sect.~\ref{sec:Conclusions}. We adopt a Hubble constant $H_0 = 73$~km~s$^{-1}$~Mpc$^{-1}$~\citep{Riess_2016} throughout the paper.

	\section{Refracted gravity}
	\label{sec:RG}
	
	Refracted gravity is a classical theory of modified gravity where dark matter is mimicked by a gravitational permittivity, $\epsilon(\rho)$, a monotonic increasing function of the local mass density $\rho$. The gravitational permittivity appears in the modified Poisson equation
	\begin{equation}
	\label{eq:PoissonRG}
	\nabla\cdot[\epsilon(\rho)\nabla\phi]=4\pi G\rho,
	\end{equation}
	where $\phi$ is the gravitational potential and $G$ the gravitational constant; the permittivity $\epsilon(\rho)$ has the asymptotic values
	\begin{equation}
	\label{eq:epsAsympt}
	\epsilon(\rho) =
	\begin{cases}
	1, & \rho \gg \rho_{\rm c}\\
	\epsilon_0, & \rho \ll \rho_{\rm c},
	\end{cases}
	\end{equation}
	where $0 < \epsilon_0 \leq 1$ is the gravitational permittivity in vacuum and $\rho_{\rm c}$ is a critical density. 
	
		Adopting a permittivity depending on the mass density introduces a spatial scale over which we need to average the mass distribution. Here, we are interested in the dynamics of galaxies and this scale has to be sufficiently smaller than the size of a galaxy. Clearly, the identification of this scale remains an open issue that needs to be addressed when RG is developed beyond its current stage of a phenomenological model that is only applied to galactic and extragalactic dynamics \citep{Matsakos_and_Diaferio_2016}.
	
	According to Eqs.~\eqref{eq:PoissonRG} and~\eqref{eq:epsAsympt}, when the density is much larger than $\rho_{\rm c}$, we recover the Newtonian Poisson equation $\nabla^2\phi=4\pi G\rho$. On the contrary, in low-density regions, the gravitational permittivity enhances the gravitational field. Refracted gravity predicts a different behaviour for the gravitational field in spherical and flattened systems. For spherical systems, the integration of Eq.~\eqref{eq:PoissonRG} yields 
	\begin{equation}
	\label{eq:Grav_field_RG_sph}
	\frac{{\rm d \phi}}{{\rm d}r} = \frac{G M( r)}{\epsilon(\rho)r^2},
	\end{equation}
	where $M(r)$ is the total mass of the system within radius $r$. The RG gravitational field has the same direction and $r$-dependence as in the Newtonian case, but it is enhanced by the inverse of the permittivity. 
	
		Equation~\eqref{eq:Grav_field_RG_sph} shows that the gravitational field is $\propto r^{-2}$ in the vacuum, where both $M(r)$ and $\epsilon(\rho)$ are constant. This behaviour of the RG gravitational field does not necessarily contradict the observed velocity dispersion profiles of elliptical galaxies~\citep{Jimenez_2013, Durazo_2017, Durazo_2018} and of GCs \citep{Hernandez_and_Jimenez_2012, Hernandez_2013, Durazo_2017, Hernandez_and_LaraDI_2020}, or the external $\rho\propto r^{-3}$ mass density profiles of spherical stellar systems \citep{Hernandez_2013b} that suggest a gravitational field falling off as $r^{-1}$. No astrophysical system is completely isolated and exactly satisfies the conditions $M(r)={\rm const}$ and $\epsilon(\rho)={\rm const}$ in its outer regions. Therefore, in principle, the permittivity $\epsilon(\rho)$ can determine an $r^{-1}$ behaviour of the gravitational field in each of these systems depending on the actual density field of its environment.
	
	For flat systems, the lines of the gravitational field are focussed towards the mid-plane of the source 
	\citep{Matsakos_and_Diaferio_2016,Cesare_2020}: the flatter the system, the stronger the redirection effect. 
	This feature explains the rotation curves of disk galaxies and might be consistent with the claimed correlation between the dark matter content of elliptical galaxies and their ellipticity \citep{Deur_2014, Deur_2020}.
	
	In this work, we solved Eq.~\eqref{eq:Grav_field_RG_sph} for three approximately spherical galaxies to determine whether the gravitational permittivity can describe the dynamics of spherical systems independently of the redirection of the field lines. In Eq.~\eqref{eq:Grav_field_RG_sph}, the mass density $\rho$ and, consequently, the cumulative mass profile $M(r)$ of each galaxy are associated with the baryonic matter alone. For the gravitational permittivity, according to~\citet{Matsakos_and_Diaferio_2016} and~\citet{Cesare_2020}, we adopted the smooth step function
	\begin{equation}
	\label{eq:eps}
	\epsilon(\rho)=\epsilon_0+(1-\epsilon_0)\frac{1}{2}\left\{\tanh\left[\text{ln}\left(\frac{\rho}{\rho_\mathrm{c}}\right)^Q\right]+1\right\}.
	\end{equation} 
	The parameter $Q$ regulates the steepness of the transition between the Newtonian and the RG regimes: the larger $Q$, the steeper the transition. With this formulation, RG has three free parameters that we expect to be universal: $\epsilon_0$, $\rho_{\rm c}$, and $Q$.

	\section{Photometric data}
	\label{sec:Photometric_data}
	
	In this section, we describe the observables that entered our mass model of each galaxy: the surface brightness of the stars, the surface number densities of the GCs populations, the mass density of the hot X-ray emitting gas, and the mass of the central supermassive black hole (SMBH). Table~\ref{tab:Quantities} lists the quantities characterising the three galaxies and the colour cut that we adopted in Sect.~\ref{sec:V_rms_GCs} to separate the GC samples into the red and blue populations.  
	
	The galaxies NGC 1407, NGC 4486, and NGC 5846 are at redshift $z<0.007$. Thus, we could neglect any $k$-correction in the photometric measures and in the distance modulus $m - M = 5 \log_{10}[D({\rm pc})] - 5$, with $m$ and $M$ the apparent and absolute magnitudes of the source. Similarly, to convert the radial coordinate $R$ projected on the sky from angular units, in arcsec, to physical units, in kpc, we adopted the relation, valid in a non-expanding Euclidean geometry,
	\begin{equation}
	\label{eq:Coversion_arcsec_kpc}
		R({\rm kpc}) = 4.84814 \times 10^{-6} D({\rm kpc}) R({\rm arcsec})\, ,
	\end{equation}
	with $D$ the distance to the galaxy. For elliptical galaxies, the radial coordinate $R$ is the circularised radius
	\begin{equation}
	\label{eq:Rcirc}
	R = \sqrt{qx'^2 + \frac{y'^2}{q}},
	\end{equation}
	where $q$ is the minor-to-major axis ratio of the galaxy on the sky, and the $x'$ and $y'$ coordinates are orientated along the galaxy major and minor axes, respectively \citep[e.g.][]{Cappellari_2008,ATLAS_XV,Napolitano_2009,Pota_2013,Pota_2015}. 
	
	\begin{table*}
		\caption{General properties of NGC 1407, NGC 4486, and NGC 5846.}
		\label{tab:Quantities}	
		\centering
		\begin{tabular}{lcccccccc}
			\hline\hline
			NGC & $D$ & $z$ & $q$ & $N_{\rm GCs}$ & $N_{\rm GCs, Blue}$ & $N_{\rm GCs, Red}$ & $V_{\rm sys}$ & $(g - i)_{\rm TH}$  \\
			& [Mpc] & & & & & & $\left[{\rm km}~{\rm s}^{-1}\right]$ & \\
			(1) & (2) & (3) & (4) & (5) & (6) & (7) & (8) & (9) \\
			\hline
			1407 & $28.05$ & $0.0068$ & $0.95$ & 379 & 153 & 148  & $1779 \pm 9$ & $0.98$ \\
			4486 & $17.2$ & $0.0042$ & $0.86$ & 737 & 480 & 199 & $1284 \pm 5$ & $0.93$ \\
			5846 & $24.2$ & $0.0059$ & $0.92$ & 195 & 91 & 102 & $1712 \pm 5$ & $0.95$ \\
			\hline
		\end{tabular}
    	\tablefoot{Column 1: galaxy name; Col. 2: distance; Col. 3: redshift; Col. 4: minor-to-major axis ratio; Col. 5: number of GCs in the initial sample; Cols. 6--7: number of blue and red GCs in the final sample (see Sects.~\ref{sec:SND_GCs} and \ref{sec:V_rms_GCs}); Col. 8: systemic velocity; Col. 9: $(g - i)$ colour threshold separating the GC population. The distance and the initial number of GCs of NGC 1407 are from~\citet{Pota_2015}, whereas the distance and the initial number of GCs of NGC 4486 and NGC 5846 are from~\citet{Pota_2013}. The number of blue and red GCs of NGC 1407 are from~\citet{Pota_2015}, whereas the number of blue and red GCs of NGC 4486 and NGC 5846 are determined from our analysis. The minor-to-major axis ratio and the systemic velocities of the three galaxies are from~\citet{Pota_2013}, as well as the $(g - i)$ colour thresholds for NGC 1407 and NGC 5846. The $(g - i)$ colour threshold for NGC 4486 is from~\citet{Strader_2011}.}
	\end{table*}

	\subsection{Stellar surface brightness profiles}
	\label{sec:SB_stars}
	
	For the surface brightness of the stars, we adopted the models derived by~\citet{Pota_2015} for NGC 1407 and by~\citet{ATLAS_XXI} for NGC 4486 and NGC 5846.
	
	\subsubsection{NGC 1407}
	\label{sec:SB_stars_NGC_1407}
	
	The stellar surface brightness profile of NGC 1407 was measured in the $B$-band with the \textit{Hubble Space telescope}/ACS~\citep{Spolaor_2008,Rusli_2013b} and in the $g$-band with the Subaru/Suprime-Cam~\citep{Pota_2013}. The $B$- and $g$-bands largely overlap; therefore,~\citet{Pota_2015} derived a unique surface brightness profile in the $B$-band by a proper transformation of the $g$-band data. NGC 1407 has ellipticity $\varepsilon = 1 - q=0.05$. \citet{Pota_2015} modelled the surface brightness profile with the Sérsic profile
	\begin{equation}
	\label{eq:Sersic}
	I_*(R) = I_{\rm e} \exp\left\{-b_{n_{\rm s}}\left[\left(\frac{R}{R_{\rm e}}\right)^\frac{1}{n_{\rm s}}-1\right]\right\},
	\end{equation}
	where 
	\begin{equation}
	\label{eq:bns}
	b_{n_{\rm s}} \approx 2n_{\rm s} - \frac{1}{3} + \frac{4}{405 n_{\rm s}} + \frac{46}{25515 n_{\rm s}^2}
	\end{equation}
	 is such that the effective radius $R_{\rm e}$ contains half of the total luminosity of the stellar distribution~\citep{Ciotti_and_Bertin_1999}, $I_{\rm e}$ is the surface brightness at radius $R_{\rm e}$, and $n_{\rm s}$ is the Sérsic index. After deconvolving for the seeing,~\citet{Pota_2015} derived $R_{\rm e} = (100 \pm 3)$~arcsec, $I_{\rm e} = 3.5 \times 10^5$~$L_{\odot,B}$~arcsec$^{-2}$, and $n_{\rm s} = 4.67 \pm 0.15$.
	
	\subsubsection{NGC 4486 and NGC 5846}
	\label{sec:SB_stars_NGC_4486_NGC_5846}
	 
	 \citet{ATLAS_XXI} measured the surface brightness of NGC 4486 and NGC 5846 in the \textit{ugriz} bands from the Sloan Digital Sky Survey DR7~\citep{Abazajian_2009} and the Wide Field Camera on the 2.5-m Isaac Newton Telescope at the Roque de los Muchachos Observatory. \citet{ATLAS_XXI} modelled the two-dimensional map of the surface brightness in the $r$-band; this band reduces the dust contamination and provides images with the optimal signal-to-noise value~\citep{ATLAS_XV}. They adopted the axisymmetric Multi-Gaussian Expansion (MGE) approach~\citep{Ems94}, which yields the surface brightness map~\citep{Cappellari_2008,ATLAS_XV}
	\begin{equation}
	\label{eq:Sigma_MGE_axi}
		I_*(x',y') = \sum_{k = 1}^{N}\frac{L_k}{2\pi\sigma_k^2q'_k}\exp\left[-\frac{1}{2\sigma^2_k}\left(x'^2 + \frac{y'^2}{q_k'^2}\right)\right],
	\end{equation}
	where $N$ is the number of Gaussian components in the MGE fit, and $L_k$, $\sigma_k$, and $0 \leq q'_k \leq 1$ are the luminosity, the standard deviation along the major axis, and the observed minor-to-major axis ratio of each Gaussian, respectively; $x'$ and $y'$ are the coordinates on the plane of the sky, where the $x'$-axis is orientated along the major axis of the galaxy. \citet{ATLAS_XXI} found $q'_k= 1$ within 5\% for almost all the Gaussians of the model of each galaxy. Indeed, NGC 4486 has ellipticity $\varepsilon = 1-q=0.14$, and NGC 5846 has $\varepsilon=0.08$. Therefore, we modelled these galaxies as spherical systems, set $q'_k=1$ for all the Gaussians, and adopted the parameters $L_k$ and $\sigma_k$ determined by~\citet{ATLAS_XXI} for Eq.~\eqref{eq:Sigma_MGE_axi} in the surface brightness profile~\citep{Cappellari_2008}
	\begin{equation}
	\label{eq:Sigma_MGE_sph}
	I_*(R) = \sum_{k= 1}^{N}\frac{L_k}{2\pi \sigma^2_k}\exp\left(-\frac{R^2}{2\sigma^2_k}\right).
	\end{equation}
	\citet{ATLAS_XXI} found $N=9$ and $N=7$ Gaussian components for NGC 4486 and NGC 5846, respectively. Tables~\ref{tab:SLUGGS_Phot_Stars_NGC_4486} and~\ref{tab:SLUGGS_Phot_Stars_NGC_5846}, reported in Appendix~\ref{sec:SB_stars_NGC_4486_NGC_5846_A}, list $L_k$ and $\sigma_k$ 
	after deconvolving for the seeing.
	Finally, the total stellar luminosity of the galaxy is~\citep{ATLAS_XV}
	\begin{equation}
	\label{eq:L_tot_MGE}
		L_{*,{\rm tot}} = \sum_{k= 1}^{N} L_k.
	\end{equation}	

    \subsubsection{Stellar 3D luminosity density profiles}
    \label{sec:nu_stars}
	
	For NGC 1407, the three-dimensional (3D) luminosity density of the stars is the Abel integral, valid in spherical symmetry,
	\begin{equation}
	\label{eq:nu}
	\nu_*(r) = -\frac{1}{\pi}\int_{r}^{+\infty} \frac{{\rm d}I_*}{{\rm d}R} \frac{{\rm d}R}{\sqrt{R^2 - r^2}}\, ,
	\end{equation}
	where $I_*(R)$ is given by Eq.~\eqref{eq:Sersic} and $r$ is the radial coordinate in three dimensions. For NGC 4486 and NGC 5846, the deprojected MGE 3D luminosity density is~\citep{Cappellari_2008}
	\begin{equation}
	\label{eq:nu_MGE}
	\nu_*(r) = \sum_{k =1}^{N}\frac{L_k}{(\sqrt{2\pi}\sigma_k)^3}\exp\left(-\frac{r^2}{2\sigma^2_k}\right).
	\end{equation}
	
	The 3D luminosity density of the stars, $\nu_*(r)$, yields the cumulative stellar luminosity profile
	\begin{equation}
	\label{eq:L_Stars}
	L_*(r) = 4\pi \int_{0}^{r} \nu_*(r^\prime) r^{\prime 2} \text{ } {\rm d}r^\prime. 
	\end{equation} 
	Setting $r=+\infty$ in Eq.~\eqref{eq:L_Stars} yields the total stellar luminosity  $L_{*,{\rm tot}}=8.53 \times 10^{10}$~$L_\odot$ for NGC 1407 \citep{Pota_2015}. Adopting  Eq.~\eqref{eq:L_tot_MGE} yields $L_{*,{\rm tot}}= 7.40 \times 10^{10}$~$L_\odot$ and $4.62 \times 10^{10}$~$L_\odot$ for NGC 4486 and NGC 5846, respectively \citep{ATLAS_XXI}.

	\subsection{Number density profiles of GCs}
	\label{sec:SND_GCs}
	
	The number density profiles of GCs were measured by \citet{Pota_2015}, \citet{Strader_2011}, and \citet{Pota_2013}, for NGC 1407, NGC 4486, and NGC 5846, respectively. The profiles were estimated separately for blue and red GCs, according to the colour thresholds listed in Table~\ref{tab:Quantities}. Below, we remind the criteria for the GC selection because we applied the same criteria in Sect.~\ref{sec:V_rms_GCs_NGC_4486_NGC_5846} to derive the kinematic profiles of the GCs in NGC 4486 and NGC 5846. 
	
	To ensure the magnitude completeness of the GCs samples in all the three galaxies, \citet{Pota_2013, Pota_2015}  and \citet{Strader_2011} removed the sources fainter than $M_i=-8.0$ in the $i$-band. This magnitude limit corresponds to the peak of the GC luminosity function, which has a roughly Gaussian shape~\citep[e.g.][]{Harris_1991,Whitmore_1995,Kundu_and_Whitmore_1998}. Moreover, the samples might be contaminated by ultra-compact dwarf galaxies (UCDs), which are a kinematically and spatially distinct population from the GCs. Therefore, \citet{Pota_2013, Pota_2015} and \citet{Strader_2011} removed the sources brighter than $M_i = -11.6$. This luminosity is 1 mag brighter than $\omega$ Cen, the brightest GC in the Milky Way, and brighter objects are likely to be UCDs. 
	Imposing the magnitude range $-11.6<M_i<-8.0$ removes 72 objects from the NGC 1407 sample, and 53 objects from the NGC 4486 sample, whereas the NGC 5846 sample remains unaffected. 
 
    In addition to this photometric selection,~\citet{Pota_2013, Pota_2015}  and \citet{Strader_2011} applied a kinematic selection on the two separated populations of blue and red GCs. Despite Galactic stars and GCs are two populations whose velocity distributions are usually well distinct, the population of stars might overlap on the low-velocity tail of the GC distribution so that a Galactic star could erroneously be identified as a GC.  Thus, \citet{Pota_2013, Pota_2015} and \citet{Strader_2011} removed the sources whose velocity is more than 3$\sigma$ discrepant from the mean velocity of the 20 closest neighbours in the GC sample, where $\sigma$ is the velocity dispersion of these 20 neighbours~\citep{Merrett_2003}. This kinematic criterion removed six, five, and two objects from the NGC 1407, NGC 4486, and NGC 5846 samples, respectively. The sizes of the final samples of blue and red GCs for each galaxy are listed in Table \ref{tab:Quantities}.
  
    \citet{Pota_2013, Pota_2015} and \citet{Strader_2011} binned each GC sample into circularised annuli, according to Eq.~\eqref{eq:Rcirc}, around the galaxy centre. The background-subtracted surface number density profiles, along with their Poissonian uncertainties, are shown in Fig.~\ref{fig:NGC_Blue_Red_NGC_1407_NGC_5846}. These uncertainties do not include possible biases due to selection effects: GC counts can miss objects either in the central region of the galaxy, because of the high surface brightness of the host galaxy, or in the galaxy outskirts, because of the star-shaped footprint of the SLUGGS survey.
    
	\subsubsection{NGC 5846}
	\label{sec:SND_GCs_NGC_5846}
	 
    We modelled the binned profiles of the surface number density of the blue and red GCs in NGC 5846  with the Sérsic profile
	\begin{equation}
	\label{eq:N_Sersic}
	N_{\rm GC}(R) = N_{\rm e}  \exp\left\{-b_{n_{\rm s}}\left[\left(\frac{R}{R_{\rm e}}\right)^\frac{1}{n_{\rm s}} - 1\right]\right\},
	\end{equation}
	where the parameters $R_{\rm e}$, $N_{\rm e}$, and $n_{\rm s}$ have similar meaning of those in Eq.~\eqref{eq:Sersic}. 
	
	We estimated the free parameters of the surface number density profiles of the blue and red GCs with a Monte Carlo Markov chain (MCMC) method with a Metropolis--Hastings acceptance criterion. Details of this algorithm are in Sect.~\ref{sec:MCMC}. We ran the MCMC for $2 \times 10^6$ steps with $10^5$ burn-in elements to achieve the convergence of the chains, according to the \citet{Geweke_1992} diagnostics. We adopted flat priors on the free parameters of the model in the ranges listed in Table~\ref{tab:SLUGGS_Phot_GCs_Priors_NGC_5846}. The upper limit of $R_{\rm e}$ is the radius of the most external data point of the surface number density; the upper limit of $N_{\rm e}$ is slightly larger than the surface number density at the smallest $R$. Therefore, the upper limits of the priors for $R_{\rm e}$ and $N_{\rm e}$ are different for the two GC populations. The distribution of the GCs, particularly the red population, generally follows the distribution of the stars of the host galaxy~\citep[e.g.][]{Brodie_and_Strader_2006}. The Sérsic index of the stellar component in giant elliptical galaxies is $\gtrsim 4$~\citep{Kormendy_2009}. We thus set to $10$ the upper limit of the prior on the Sérsic index for the two GCs sub-populations.
	
	Our posterior distributions are single-peaked. Thus, we adopted their medians as our parameter estimates and the range between the 15.9 and 84.1 percentiles, which includes 68\% of the posterior cumulative distribution, as our parameter uncertainties. These parameters are listed in Table~\ref{tab:SLUGGS_Phot_GCs_NGC_1407_NGC_4486_NGC_5846}. They provide the curves shown in the right panel of Fig.~\ref{fig:NGC_Blue_Red_NGC_1407_NGC_5846}.
	
	\subsubsection{NGC 1407 and NGC 4486}
	\label{sec:SND_GCs_NGC_1407_and_NGC_4486}
	
\begin{figure*}
    	\centering
    	\includegraphics[width=17cm]{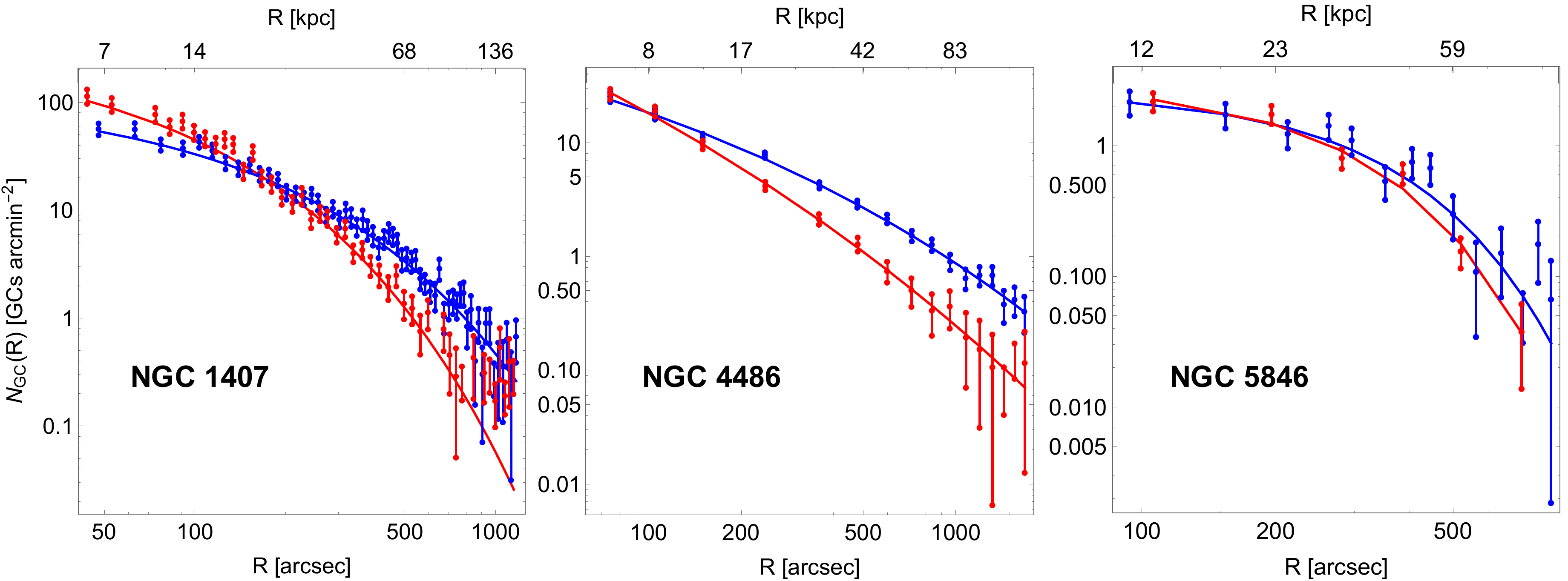}
    	\caption{Surface number density profiles of blue (blue symbols and blue lines) and red (red symbols and red lines) GCs. Blue and red dots with error bars show the binned profiles; the solid lines are the models. }
    	\label{fig:NGC_Blue_Red_NGC_1407_NGC_5846}
    \end{figure*}

	For NGC 1407 and NGC 4486, we adopted the models available in the literature. \citet{Pota_2015} modelled their data of NGC 1407 with Eq.~\eqref{eq:N_Sersic}. For NGC 4486, \citet{Strader_2011} adopted the slightly different parametrisation 
	\begin{equation}
	\label{eq:N_Sersic_Strader_2011}
		N_{\rm GC}(R) = N_0  \exp\left[-\left(\frac{R}{R_{\rm s}}\right)^\frac{1}{m}\right].
	\end{equation} 
	\citet{Pota_2015} and \citet{Strader_2011} derived the parameters listed in Table~\ref{tab:SLUGGS_Phot_GCs_NGC_1407_NGC_4486_NGC_5846} that yield the curves shown in Fig.~\ref{fig:NGC_Blue_Red_NGC_1407_NGC_5846}.
	
	In Table~\ref{tab:SLUGGS_Phot_GCs_NGC_1407_NGC_4486_NGC_5846}, we also list the $\chi^2$ of these models, for comparing the goodness of the fits of the six GC samples. The 3D number density of each GC sub-sample is 
	\begin{equation}
	\label{eq:nu_Sersic}
	\nu_{\rm GC}(r) = -\frac{1}{\pi}\int_{r}^{+\infty}\frac{{\rm d}N_{\rm GC}}{{\rm d}R}\frac{{\rm d}R}{\sqrt{R^2 - r^2}}\, .
	\end{equation}

	\begin{table*}
		\caption{Priors of the parameters of the GC surface number density models of NGC 5846.}
		\label{tab:SLUGGS_Phot_GCs_Priors_NGC_5846}	
		\centering
		\begin{tabular}{lccc}
			\hline\hline
			GC sample & $N_{\rm e}$ & $R_{\rm e}$ & $n_{\rm s}$ \\
			& $\left[\frac{\rm GCs}{{\rm arcmin}^2}\right]$ & [arcsec] & \\
			(1) & (2) & (3) & (4) \\
			\hline
			Blue & $U(0.00,2.24]$ & $U(0.00,833]$ & $U(0.00,10.0]$  \\
			Red  & $U(0.00,2.55]$ & $U(0.00,714]$ & $U(0.00,10.0]$  \\
			\hline
		\end{tabular}
	\tablefoot{The symbol $U$ stands for uniform distribution.}
	\end{table*}

    \begin{table*}
    	\caption{Parameters of the models of the surface number density of the GCs.}
    	\label{tab:SLUGGS_Phot_GCs_NGC_1407_NGC_4486_NGC_5846}	
    	\centering
    	\begin{tabular}{lccccc}
    		\hline\hline
    		NGC & GC sample & $N_{\rm e},\, N_0$ & $R_{\rm e},\, R_{\rm s}$  & $n_{\rm s},\, m$ & $\chi^2_{{\rm red},\nu}$\\
    		&&$\left[\frac{\rm GCs}{{\rm arcmin}^2}\right]$ & [arcsec] && \\
    		(1) & (2) & (3) & (4) & (5) & (6) \\
    		\hline
    		1407 & Blue & $7 \pm 1 $ & $346 \pm 29$ & $1.6 \pm 0.2$ &  $0.90$ \\
    		     & Red & $20 \pm 2$ & $169 \pm 7$ & $1.6 \pm 0.2$ &  $1.29$ \\
    		4486 & Blue & $(6.1 \pm 2.9) \times 10^2 $ & $1 \pm 2$ & $3.69 \pm 0.47$ & $0.54$ \\
    		     & Red & $(5.1 \pm 1.6) \times 10^4$ & $(1.6 \pm 1.7) \times 10^{-3}$ & $5.33 \pm 0.24$ &  $0.28$ \\
    		5846 & Blue &$0.87^{+0.16}_{-0.17}$ & $309^{+27}_{-22}$ & $0.71^{+0.27}_{-0.18}$ & $1.06$ \\
    		     & Red & $1.00^{+0.15}_{-0.15}$ & $266^{+18}_{-17}$ & $0.72^{+0.23}_{-0.16}$ & $1.21$ \\
    		\hline
    	\end{tabular}
    	\tablefoot{Column 1: galaxy name; Col. 2: colour of the GC population; Cols. 3--5: parameters of the model of the surface number density; Col. 6: reduced chi-square, $\chi^2_{{\rm red},\nu}$, for $\nu$ degrees of freedom. The parameters $\{N_{\rm e},R_{\rm e},n_{\rm s}\}$, adopted for NGC 1407 and NGC 5846, refer to Eq.~\eqref{eq:N_Sersic} and are from \citet{Pota_2015} and our MCMC analysis, respectively; the parameters $\{N_0,R_{\rm s},m\}$, adopted for NGC 4486, refer to Eq.~\eqref{eq:N_Sersic_Strader_2011} and are from \citet{Strader_2011}.}
    \end{table*}
    
    \subsection{Mass density profiles of the X-ray gas}
    \label{sec:rho_gas}
    
    For the 3D mass density profile of the X-ray emitting gas, we adopted results from the literature for NGC 1407~\citep{Zhang_2007} and NGC 4486~\citep{Fabricant_1980}. For NGC 5846, we derived our model of the density profile from the estimate of~\citet{Paggi_2017} of a discrete profile.
    
    \subsubsection{NGC 1407}
    \label{sec:rho_gas_NGC_1407}
     
    \citet{Zhang_2007} assumed spherical symmetry and modelled the 3D number density profile of the gas, $n_{\rm g}(r)$, with the two-$\beta$ function
    \begin{equation}
    \label{eq:Two_beta}
    	n_{\rm g}^2(r) = n_{\rm g,1}^2 \left[1 + \left(\frac{r}{R_{\rm c,1}}\right)^2 \right]^{-3\beta_1} + n_{\rm g,2}^2 \left[1 + \left(\frac{r}{R_{\rm c,2}}\right)^2 \right]^{-3\beta_2}. 
    \end{equation}
    \citet{Zhang_2007} derived the parameters $n_{\rm g, 1} = 0.1$~cm$^{-3}$, $n_{\rm g, 2} = 3.65 \times 10^{-3}$~cm$^{-3}$, $R_{\rm c, 1} = (8.42 \pm 0.70)$~arcsec, $R_{\rm c, 2} = (58.3 \pm 0.7)$~arcsec, $\beta_1 = 0.70 \pm 0.01$, and $\beta_2 = 0.45 \pm 0.01$ by fitting the X-ray surface brightness profile measured with \textit{Chandra} ACIS in the 0.7--7~keV band and with \textit{ROSAT} PSPC in the 0.2--2~keV band. \citet{Zhang_2007} reported $n_{\rm g, 1}$ and $n_{\rm g, 2}$ without uncertainty.   
    
    The 3D mass density profile of the gas entering our model is
    \begin{equation}
    \label{eq:ng_to_rhog}
    	\rho_{\rm g}(r) = \mu m_{\rm H} n_{\rm g}(r)\, ,
    \end{equation}
    where $m_{\rm H} = 1.66054 \times 10^{-27}$~kg is the atomic unit mass and $\mu = 0.6$ is the mean molecular weight, which assumes a completely ionised gas and a chemical composition with zero metallicity. The normalisations $n_{\rm g,1}$ and $n_{\rm g,2}$ yield $\rho_{\rm g,1} = 9.96 \times 10^{-26}$~g~cm$^{-3}$ and $\rho_{\rm g,2} = 3.64 \times 10^{-27}$~g~cm$^{-3}$.
    
    \subsubsection{NGC 4486}
    \label{sec:rho_gas_NGC_4486}
    
        \citet{Fabricant_1980} modelled the 3D mass density profile of the gas with the function
    	\begin{equation}
    	\label{eq:rhor_gas_NGC_4486}
    	\rho(r) = \frac{\rho_0}{(1 + b'r^2 + c'r^4 + d'r^6)^{n'}}\, ,
    	\end{equation}
    	obtained by deprojecting the model for the X-ray surface brightness measured in the 0.7--3.0~keV energy band with the \textit{Einstein Observatory}. \citet{Fabricant_1980} provided the model parameters without uncertainty: $b' = 9.724 \times 10^{-1}$~arcmin$^{-2}$, $c' = 3.810 \times 10^{-3}$~arcmin$^{-4}$, $d' = 2.753 \times 10^{-8}$~arcmin$^{-6}$, $n' = 0.59$, and $\rho_0 = 1.0 \times 10^{-25}$~g~cm$^{-3}$.

    \subsubsection{NGC 5846}
    \label{sec:rho_gas_NGC_5846}
    
\begin{figure}
    	\resizebox{\hsize}{!}{\includegraphics{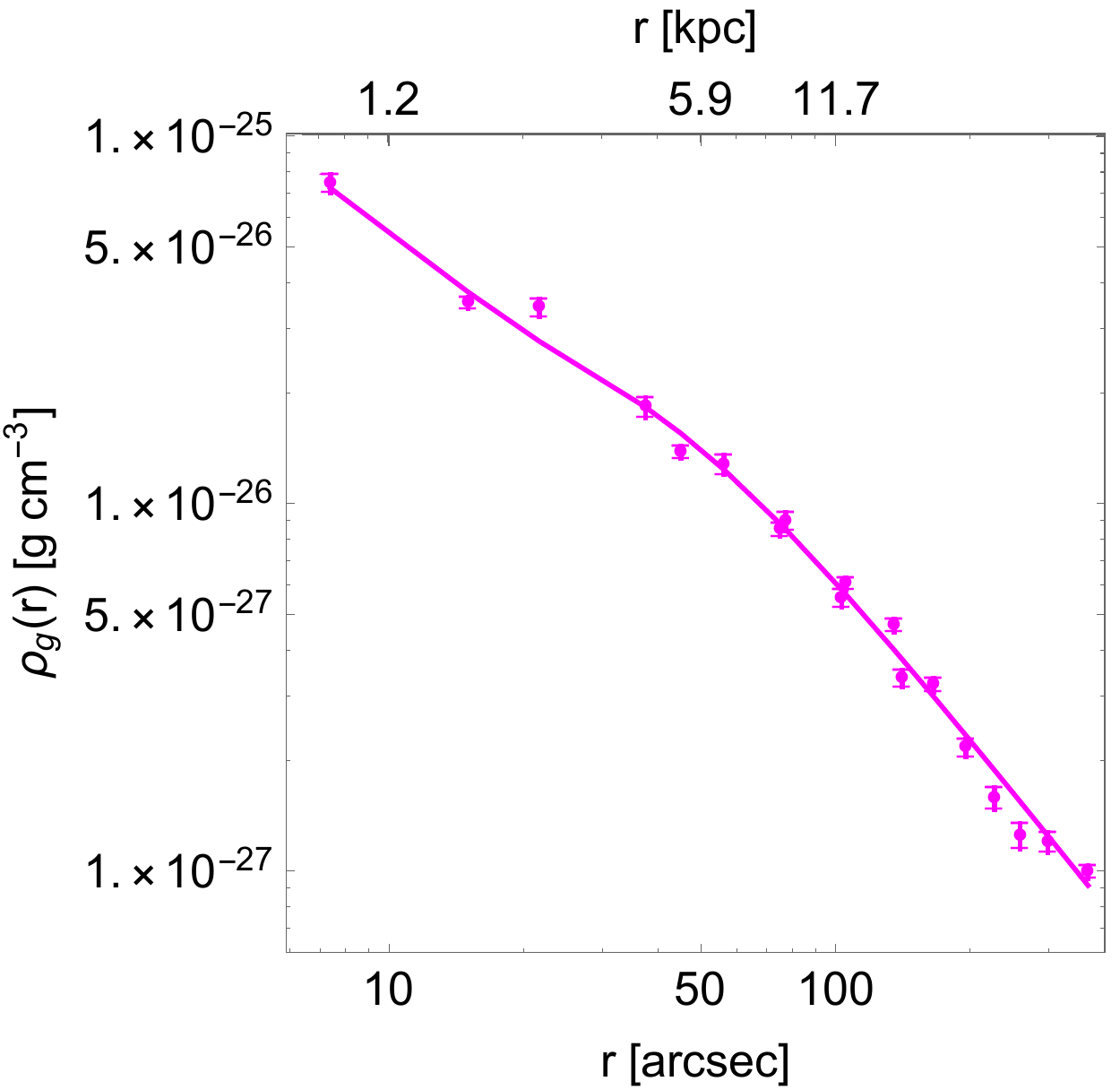}}
    	\caption{Three-dimensional mass density profile of the hot gas of NGC 5846. The solid circles with error bars show the discrete profile derived by \citet{Paggi_2017}; the curve shows our model.
    	}
    	\label{fig:rho_gas_goncm3_NGC_5846}
    \end{figure}

     We derived the continuous 3D gas mass density profile from the discrete mass density profile estimated by~\citet{Paggi_2017}. \citet{Paggi_2017} used the X-ray surface brightness measurements obtained with the \textit{Chandra} ACIS and \textit{XMM-Newton} MOS instruments. We modelled this discrete profile with the two-$\beta$ function of Eq.~\eqref{eq:Two_beta}, where the two normalisation constants are now $\rho_{\rm g,1}$ and $\rho_{\rm g,2}$. We estimated the parameters of the gas mass density with the MCMC method described in Sect.~\ref{sec:MCMC} adopting the Metropolis-Hasting acceptance criterion. We ran the chains for $2 \times 10^6$ steps with $10^5$ burn-in elements. To test the chain convergence, we adopted the~\citet{Geweke_1992} diagnostics. We adopted uniform priors in the intervals listed in Table~\ref{tab:SLUGGS_rho_gas_Priors_Parameters_merged_NGC_5846}. The upper limits of $R_{\rm c,1}$ and $R_{\rm c,2}$ are the radii of the most external data point of the measured gas mass density profile. The upper limits of $\rho_{\rm g,1}$ and $\rho_{\rm g,2}$ are slightly larger than the data point with the smallest $R$. Our posterior distributions are single-peaked, and thus we adopted the medians of the posterior distributions as our parameter estimates and the range between the 15.9 and 84.1 percentiles as our parameter uncertainties. The parameters of the model, along with their uncertainties, are listed in Table~\ref{tab:SLUGGS_rho_gas_Priors_Parameters_merged_NGC_5846}. Figure~\ref{fig:rho_gas_goncm3_NGC_5846} shows the model superimposed on the data.

    \begin{table*}
    	\caption{Priors and parameters of the 3D mass density model of the hot X-ray emitting gas of NGC 5846.}
    	\label{tab:SLUGGS_rho_gas_Priors_Parameters_merged_NGC_5846}	
    	\centering
    	\begin{tabular}{lccccc}
    		\hline\hline
    		$\rho_{\rm g,1}$ & $\rho_{\rm g,2}$  & $R_{\rm c,1}$ & $R_{\rm c,2}$ & $\beta_1$ & $\beta_2$\\
    		$\left[10^{-26}\frac{\rm g}{{\rm cm}^3}\right]$ & $\left[10^{-26}\frac{\rm g}{{\rm cm}^3}\right]$ & [arcsec] & [arcsec] & & \\
    		(1) & (2) & (3) & (4) & (5) & (6) \\
    		\hline
    		$U(0,21.0]$ & $U(0,21.0]$ & $U(0,368]$ & $U(0,368]$ & $U(0,3]$ & $U(0,3]$ \\
    		$11.0^{+4.2}_{-1.3}$ & $3.15^{+0.46}_{-0.49}$ & $11^{+1}_{-1}$ & $36^{+4}_{-5}$ & $0.93^{+0.32}_{-0.18}$ & $0.507^{+0.008}_{-0.015}$ \\
    		\hline
    	\end{tabular}
    \tablefoot{The symbol $U$ stands for uniform distribution.}
    \end{table*}
    
    \subsection{SMBH mass}
    \label{sec:M_SMBH}
    
    We included the contribution of the central SMBH in our mass model. We modelled each SMBH as a point mass, adopting the SMBH masses provided by~\citet{Rusli_2013b}: $M_\bullet = 4.5^{+0.9}_{-0.4}\times 10^9$~$M_\odot$, $ 6.2^{+0.4}_{-0.5}\times 10^9$~$M_\odot$, and $ 1.1^{+0.1}_{-0.1}\times 10^9$~$M_\odot$ for NGC 1407, NGC 4486, and NGC 5846, respectively.

	\section{Spectroscopic data}
	\label{sec:Spectroscopic_data}
	
	We used the photometric information described in the previous section to build our model of the mass distribution for each galaxy. We obtained our dynamical model by combining this information with the spectroscopic information on the velocity dispersion profiles of the stars (Sect.~\ref{sec:V_rms_stars}) and the GCs (Sect.~\ref{sec:V_rms_GCs}).
	 
	\subsection{Stellar velocity dispersion profiles}
	\label{sec:V_rms_stars}
	
	For NGC 1407, we used the velocity dispersion profile provided by \citet{Pota_2015}. For NGC 4486 and NGC 5846, we derived the velocity dispersion profiles from the two-dimensional data of the ATLAS$^{\rm 3D}$ survey \citep{ATLAS_I}.
	
	\subsubsection{NGC 1407}
	\label{sec:V_rms_stars_NGC_1407}
	
	\citet{Pota_2015} derived the root-mean-square velocity dispersion profile from two different data sets, depending on the radial range. For radii in the range  $[0, 30]$~arcsec, they used the long-slit data, along the major axis of the galaxy, from the European Southern Observatory Faint Object Spectrograph and Camera (v.2) (EFOSC2)~\citep{Proctor_2009}. For radii in the range  $[40, 110]$~arcsec, they used the multislit Keck/DEIMOS data~\citep{Foster_2016}, again along the galaxy major axis. In the overlapping range  $[30, 40]$~arcsec, they kept both data sets.
	
	The root-mean-square velocity dispersion profile of the stars at the circularised projected radius $R$ is
	\begin{equation}
	\label{eq:vrms_Stars_general}
	V_{\rm rms}(R) = \left[V_{\rm rot}^2(R) + \sigma^2(R)\right]^{1/2}\, ,
	\end{equation}
	where $\sigma(R)$ is the stellar velocity dispersion and $V_{\rm rot}(R)= V(R) - V_{\rm sys}$ is the major axis stellar rotation velocity in the galaxy frame,  
	with $V(R)$ the observed rotation velocity, and $V_{\rm sys}$ the galaxy systemic velocity. 
		The analysis of~\citet{Proctor_2009} confirms that NGC 1407 is a slow rotator, namely $V_{\rm rot} \ll \sigma$; therefore,  Eq.~\eqref{eq:vrms_Stars_general} yields $V_{\rm rms}(R) \thickapprox \sigma(R)$.
	
	In our dynamical model described in Sect.~\ref{sec:Mass_modelling}, we assumed an orbital anisotropy parameter $\beta$ independent of the 3D radius $r$, as~\citet{Pota_2015} did. However, by fitting Schwarzschild orbit-superposition models to their kinematic data,~\citet{Thomas_2014} inferred a variable $\beta(r)$ within $R=2$~arcsec of NGC 1407. Therefore,~\citet{Pota_2015} only considered the $V_{\rm rms}$ profile beyond  $R = 2$~arcsec = $0.272$~kpc. The top-left panel of Fig.~\ref{fig:vrms_Data_NGC_1407_NGC_4486_NGC_5846} shows the $V_{\rm rms}(R)$ profile of the stars obtained by~\citet{Pota_2015}.
	
	\subsubsection{NGC 4486 and NGC 5846}
	\label{sec:V_rms_stars_NGC_4486_NGC_5846}
	
	The data publicly available on the ATLAS$^{3{\rm D}}$ website\footnote{\url{http://www-astro.physics.ox.ac.uk/atlas3d/}} provide the two-dimensional map of $V$ and $\sigma$ of the galaxy on the plane of the sky, where $V$ and $\sigma$ are the mean line-of-sight velocity of the stars and their velocity dispersion. As $V$ and $\sigma$, we adopted the values estimated from the simple Gaussian parametrisation, rather than from the Gauss-Hermite parametrisation, of the velocity distribution of the stars derived from the spectral line profile~\citep{Cappellari_2008,ATLAS_XV}. In NGC 4486 and NGC 5846, 99.6\% and 97.9\% of the data points, respectively, have $V/\sigma \leq 0.2$. Therefore these galaxies, like NGC 1407, are slow rotators, and we have $V_{\rm rms} \thickapprox \sigma$.
	
	From the map, we iteratively removed the values of $V_{\rm rms}$ that deviate more than 3 standard deviations from the mean $V_{\rm rms}$ of the entire map~\citep{ATLAS_XV}. For NGC 4486 and NGC 5846, the procedure converges after seven and six iterations and removes 155 and 201 data points, respectively. This removal is necessary because some data points could be spurious, because of the presence of Milky Way stars or problematic bins at the edge of the field of view~\citep{ATLAS_XV}. To fold and average the data with respect to the galaxy centre and thus transform the two-dimensional map $V_{\rm rms}$ into a one-dimensional profile, $V_{\rm rms}(R)$, we rotated the galaxy map counter-clockwise from the celestial coordinates to the  system of coordinates $(x^\prime,y^\prime)$ of Eq.~\eqref{eq:Rcirc}, such that $x^\prime$ is aligned with the photometric major axis of the galaxy~\citep{ATLAS_II}. 
	
\begin{figure*}
		\centering
		\includegraphics[width=17cm]{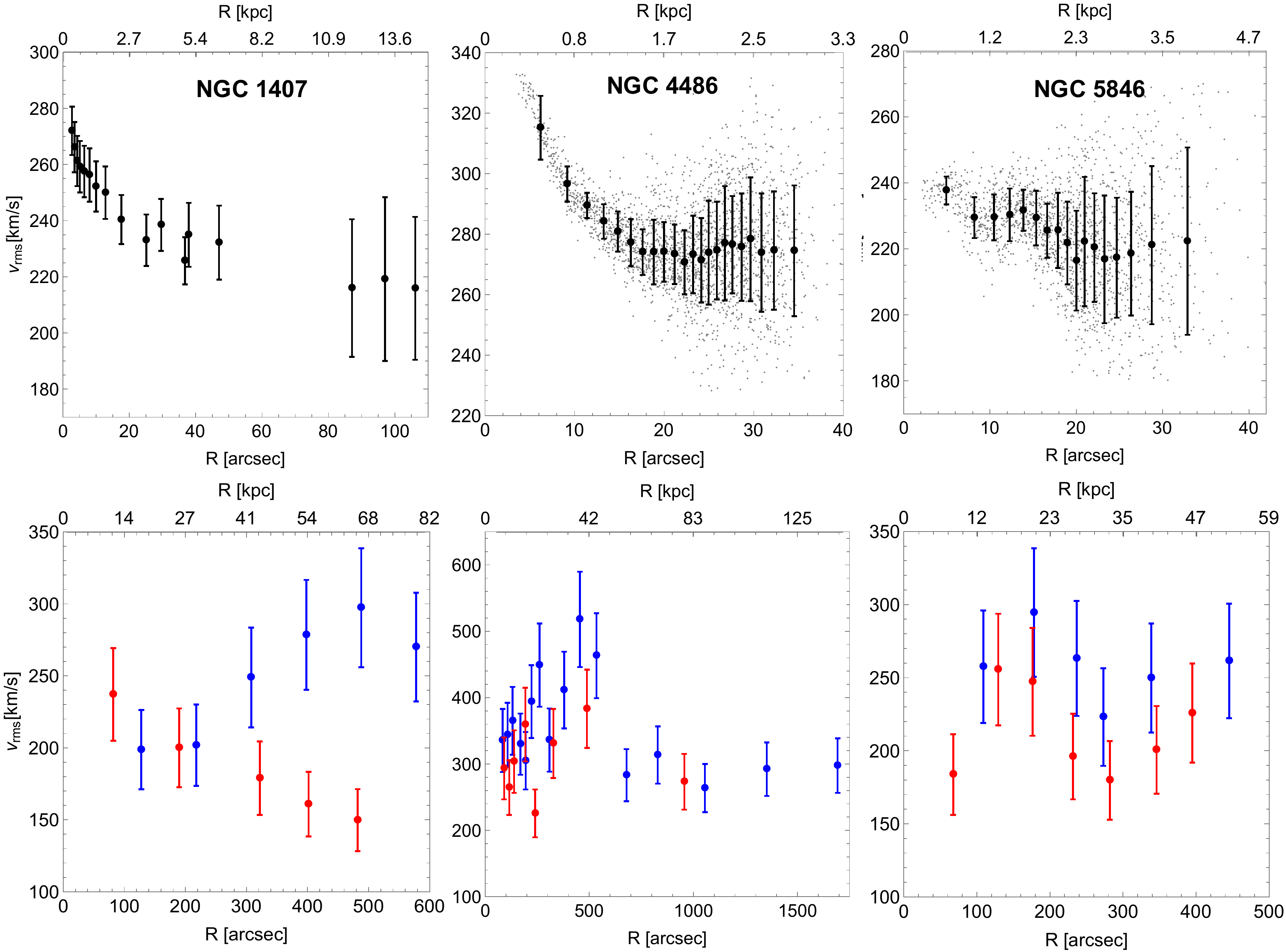}
		\caption{Root-mean-square velocity dispersion profiles of the stars (\textit{upper panels}) and of the GCs (\textit{lower panels}). Each column refers to the galaxy indicated in the \textit{upper panels}. \textit{Upper middle and right panels}: grey dots and black dots with error bars show the unbinned and binned profiles, respectively. \textit{Lower panels}: red and blue dots with error bars refer to the blue and red GC populations, respectively. We note the different radial and velocity ranges of the panels.}
		\label{fig:vrms_Data_NGC_1407_NGC_4486_NGC_5846}
	\end{figure*}

	This procedure yields $V_{\rm rms}(R)$ profiles with 2444 and 1752 data points for NGC 4486 and NGC 5846, respectively. These profiles are shown in the upper middle and right panels of Fig.~\ref{fig:vrms_Data_NGC_1407_NGC_4486_NGC_5846}. These numbers are $\sim$2 orders of magnitude larger than the number of data points in the kinematic profiles of the GC populations of the two galaxies (see Sect.~\ref{sec:V_rms_GCs_NGC_4486_NGC_5846}). As confirmed by preliminary tests, with these unbalanced data sets the dynamical models are driven by the $V_{\rm rms}(R)$ of the stars and are basically insensitive to the GC kinematic profiles. To overcome this problem, we binned the stellar kinematic data. Each bin contains a constant number of data points: $N=111$ for NGC 4486 and $N=103$ for NGC 5846. NGC 4486 and NGC 5846 have now 22 and 17 binned data, respectively. These numbers of binned data are now comparable to the numbers of bins adopted for the blue and red GCs profiles (see Sect.~\ref{sec:V_rms_GCs_NGC_4486_NGC_5846}) and guarantee that the stars and GCs data have a comparable weight in constraining our dynamical model. The values associated to the 22 and 17 binned stellar data are the medians, both in the $V_{\rm rms}$ and $R$ directions, of the values of the $N$ data points in the bin. The uncertainties are the semi-interval between the 15.9 and the 84.1 percentiles of the distributions, which are nearly symmetric, of all the data points in the bin. The binned data are shown by the black dots with error bars in the upper middle and right panels of Fig.~\ref{fig:vrms_Data_NGC_1407_NGC_4486_NGC_5846}.
    
	In general, massive early-type galaxies, such as these two galaxies, show negative orbital anisotropy parameters in their central regions, indicating tangential orbits; on the contrary, in the outer regions, the orbits appear radial~\citep{Thomas_2014,Rantala_2019}. Numerical simulations show that this difference in the stellar orbits  might originate if massive elliptical galaxies form from the merging of two progenitors with mass ratios larger than $1/3$ \citep{Rantala_2019}: at the centre of the new-born galaxy, the gravitational torques caused by the merging of the SMBHs of the two parent galaxies cause a reversal of the orbit behaviour, from radial to tangential. Therefore, because we assumed below an orbital anisotropy parameter $\beta$ independent of radius $r$, we excluded the innermost 2 arcsec in the kinematic profiles of NGC 4486 and NGC 5846, consistently with the prescription adopted for NGC 1407 by \citet{Pota_2015} (Sect.~\ref{sec:V_rms_stars_NGC_1407}). This projected radius of 2 arcsec corresponds to $0.167$~kpc and $0.235$~kpc for NGC 4486 and NGC 5846, respectively, and thus is comparable to the radius of  $0.272$~kpc of NGC 1407.
	
	\subsection{Velocity dispersion profiles of the GCs}
	\label{sec:V_rms_GCs}
	
	For NGC 4486 and NGC 5846, the velocity dispersion profiles of the two populations of GCs are not available in the literature. Therefore, we derived these profiles from the GC catalogues of~\citet{Strader_2011} and \citet{Pota_2013} for NGC 4486 and NGC 5846, respectively. For the velocity dispersions of the GCs in NGC 1407, we adopted the profiles estimated by \citet{Pota_2015}.

	\subsubsection{NGC 4486 and NGC 5846}
	\label{sec:V_rms_GCs_NGC_4486_NGC_5846}
	
	To derive the $V_{\rm rms}(R)$ profiles of the GCs in these two galaxies, we separated the GC sample into the blue and red sub-populations, according to the colour thresholds listed in Table~\ref{tab:Quantities}. To each sub-sample, we applied the photometric and kinematic selection criteria described in Sect.~\ref{sec:SND_GCs} to recover the same sub-samples used to derive the number density profiles.
 
	We binned each GC population into circularised annuli centred on the galaxy centre. Each annulus contains the same number of GCs. Specifically, each bin contains $N=30$ and $N=25$ GCs for the blue and red populations in NGC 4486 and $N=26$ GCs for both populations in NGC 5846. These numbers are a trade-off between bins that are too poor and profiles that are too coarse. For NGC 4486, we obtained 16 and 8 bins for the blue and the red GCs, respectively. For NGC 5846, we obtained six and seven bins for the blue and the red GCs, respectively. In NGC 5846, some GCs are common to contiguous bins, so that we have a sufficient number of bins despite the small total number of GCs. The radius $R$ of each bin is the median of the radial coordinates of the GCs in the bin. 
	
	The $V_{\rm rms}$ of the GCs in each bin is~\citep{Pota_2013,Pota_2015}
	\begin{equation}
	\label{eq:Vrms_GCs}
	V_{\rm rms}^2 = \frac{1}{N}\sum_{i = 1}^{N}[(V_{{\rm rad},i} - V_{\rm sys})^2 - (\Delta V_{{\rm rad},i})^2]\, ,
	\end{equation} 
	where $V_{{\rm rad},i}$ is the radial velocity of the $i$-th GC in the radial bin and $\Delta V_{{\rm rad},i}$ is its uncertainty. 
	We took $V_{\rm rad}$ and $\Delta V_{\rm rad}$ from~\citet{Strader_2011}, for NGC 4486, and~\citet{Pota_2013}, for NGC 5846. To estimate the errors on each value of $V_{\rm rms}$, we used the procedure reported in~\citet{Danese_1980}, as performed by~\citet{Pota_2015}. 
	For convenience, we report this procedure in Appendix~\ref{sec:Errors_Vrms_data_GCs}. 
	
	The kinematic profiles of the GCs are shown in Fig.~\ref{fig:vrms_Data_NGC_1407_NGC_4486_NGC_5846}. These profiles are $\sim$48 and $\sim$13 times more extended than the kinematic profiles of the stars for NGC 4486 and NGC 5846, respectively. In NGC 4486, the $V_{\rm rms}(R)$ of the GC populations show a bell-like shape that peaks at large radii, specifically at $\sim$500~arcsec $=42$~kpc. This feature suggests that the GC kinematics is influenced by the gravitational potential well of the Virgo cluster whose central giant elliptical galaxy is exactly NGC 4486 \citep{Strader_2011}. This shape is common to the stellar velocity dispersion profiles of other elliptical galaxies located at the centre of galaxy clusters, such as the brightest cluster galaxy of Abell 383~\citep{Geller_2014}.

	\subsubsection{NGC 1407}
	\label{sec:V_rms_GCs_NGC_1407}
	
	\citet{Pota_2015} derived the GC velocity dispersion profiles with the same procedure that we adopted here for NGC 4486 and NGC 5846. 
	\citet{Pota_2015} obtained their kinematic data from the Keck/DEep Imaging Multi-Object Spectrograph (DEIMOS) instrument. 
	The $V_{\rm rms}(R)$ profiles of the two GC samples are based on 153 blue GCs and 148 red GCs (see Table~\ref{tab:Quantities}), and they are shown in Fig.~\ref{fig:vrms_Data_NGC_1407_NGC_4486_NGC_5846}. They are $\sim$5 times more extended than the star kinematic profile.

	\section{Dynamical model}
	\label{sec:Mass_modelling}
	
	We now describe our dynamical model (Sect.~\ref{sec:Mass_modelling_Basic_equations}) and the MCMC approach we adopted to estimate our model parameters (Sect.~\ref{sec:MCMC}).
	
	\subsection{Basic equations}
	\label{sec:Mass_modelling_Basic_equations}
	
	We modelled the kinematics of the three E0 galaxies by assuming spherical symmetry and no net rotation, as suggested by the observed slow rotation of the galaxies. We modelled the velocity dispersion profile of each dynamical tracer ${\rm t}=\{*, \mathrm{R}, \mathrm{B}\}$, namely stars, red GCs, and blue GCs, as
	\begin{equation}
	\label{eq:Vrms_Model}
	V_{\rm rms, t}^2(R) = \frac{2}{I_{\rm t}(R)} \int_{R}^{+\infty} K\left(\beta_{\rm t}, \frac{r}{R}\right) \nu_{\rm t}(r)  \frac{{\rm d} \phi}{{\rm d} r} r\text{ }{\rm d}r\, ,
	\end{equation}
    which is the solution to the spherical Jeans equations~\citep{Jeans_1915,Cappellari_2008,Mamon_and_Lokas_2005,Pota_2015}.
	In Eq.~\eqref{eq:Vrms_Model}, $R$ is the circularised radius projected onto the sky according to Eq.~\eqref{eq:Rcirc}, whereas $r$ is the 3D radius; $I_{\rm t}(R)$ is the surface brightness of the stars or the surface number density of GCs; $\nu_{\rm t}(r)$ is the 3D luminosity density of the stars or the 3D number density of GCs; $\phi(r)$ is the gravitational potential; and $\beta_{\rm t} = 1 - \sigma^2_\theta/\sigma^2_r$ is the orbital anisotropy parameter, with $\sigma_\theta$ and $\sigma_r$ the velocity dispersions in the tangential and the radial directions, respectively. We assumed $\beta_{\rm t}$ to be independent of $r$, as discussed at the end of Sects.~\ref{sec:V_rms_stars_NGC_1407} and~\ref{sec:V_rms_stars_NGC_4486_NGC_5846}. $K$ is the kernel  
	\begin{equation}
	\label{eq:Kernel_beta}
	\begin{split}
	K\left(\beta_{\rm t}, \frac{r}{R}\right) = &\frac{1}{2}\left(\frac{r}{R}\right)^{2\beta_{\rm t}-1}\bigg[\left(\frac{3}{2}-\beta_{\rm t}\right)\sqrt{\pi}\frac{\Gamma(\beta_{\rm t}-\frac{1}{2})}{\Gamma(\beta_{\rm t})}\\
	&+\beta_{\rm t} B_{\frac{R^2}{r^2}}\left(\beta_{\rm t} + \frac{1}{2}, \frac{1}{2}\right) - B_{\frac{R^2}{r^2}}\left(\beta_{\rm t} - \frac{1}{2}, \frac{1}{2}\right) \bigg], \\
	\end{split}
	\end{equation}
	(see Eq. (A16) in~\citealt{Mamon_and_Lokas_2005}) with $\Gamma(z)$ the Euler $\Gamma$ function and $B_x(a, b)$ the incomplete beta function.
	By inserting the RG gravitational field, Eq.~\eqref{eq:Grav_field_RG_sph}, into Eq.~\eqref{eq:Vrms_Model}, we obtain
	\begin{equation}
	\label{eq:Vrms_Model_RG}
	V_{\rm rms, t}^2(R) = \frac{2G}{I_{\rm t}(R)} \int_{R}^{+\infty} K\left(\beta_{\rm t}, \frac{r}{R}\right) \nu_{\rm t}(r) \frac{M(r)}{\epsilon(\rho)} \frac{{\rm d}r}{r} \, .
	\end{equation}
	
\begin{figure*}
		\centering
		\includegraphics[width=17cm]{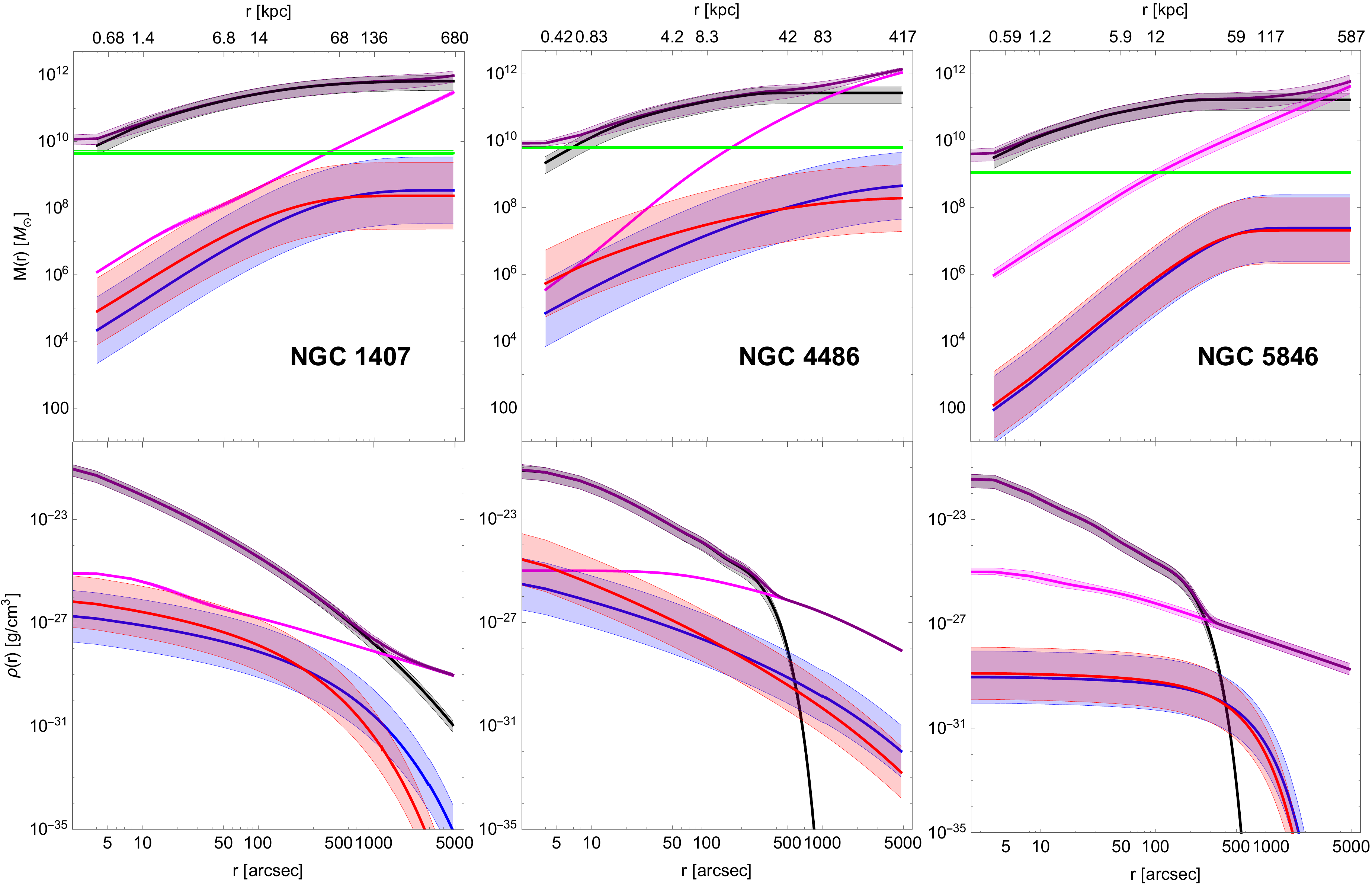}
		\caption{Cumulative mass (\textit{upper panels}) and mass density (\textit{lower panels}) profiles of the baryonic components of each galaxy: stars (black solid line), blue GCs (blue solid line), red GCs (red solid line), and hot X-ray emitting gas (magenta solid line); the green solid lines in the \textit{upper panels} show the SMBH contributions. The purple solid lines show the sum of all the baryonic contributions. The stellar profiles assume $\Upsilon=7.6$~$M_\odot/L_\odot$ for NGC 1407, and $\Upsilon=3.6$~$M_\odot/L_\odot$ for NGC 4486 and NGC 5846. The grey shaded area around the stellar profiles shows the mass variation by adopting mass-to-light ratios in the range   $[4.0,11.2]$~$M_\odot/L_\odot$, for NGC 1407, and  $[1.7,5.5]$~$M_\odot/L_\odot$, for NGC 4486 and NGC 5846, in the $B$- and $r$-band, respectively. The profiles of blue and red GCs assume a GC mass $M_{\rm GC}=10^5$~$M_\odot$; the blue and red shaded areas show the mass variation by adopting $M_{\rm GC}$ in the range $[10^4,10^6]$~$M_\odot$. The magenta shaded areas show the uncertainties on the adopted gas mass density profiles; the gas profiles of NGC 4486 do not have a shaded area because the uncertainties on the profiles are unavailable. The green solid lines and shaded areas in the \textit{upper panels} are the masses of the SMBH and their uncertainties. The purple shaded areas show the possible range of the total baryonic profiles. Each column refers to the galaxy indicated in the \textit{upper panels}.}
		\label{fig:Mass_profiles}
	\end{figure*}

	In our model, $M(r)$ is the baryonic mass alone, namely 
	\begin{equation}
	\label{eq:Total_M_profile}
		M(r) = M_*(r) + M_{\rm g}(r) + M_\bullet(r)\, ,
	\end{equation}
	where $M_*(r)$, $M_{\rm g}(r)$, and $M_\bullet(r)$ are the cumulative mass profiles of the stars, X-ray emitting gas, and SMBH, respectively.  We verified that the contribution of the GCs to the total galaxy mass profile is smaller than 0.5\% at all radii (upper panels of Fig.~\ref{fig:Mass_profiles}) and that the maximal contribution of the GCs to the total mass density profile is $\sim$$1$\% (lower panels of Fig.~\ref{fig:Mass_profiles}). We therefore ignored the contribution of the GCs to the mass profile of Eq.~\eqref{eq:Total_M_profile}. We also ignored the GC contribution to the density profile 
	\begin{equation}
	\label{eq:Total_rho_profile}
	\rho(r) = \rho_*(r) + \rho_{\rm g}(r) + \rho_\bullet(r)\, ,
	\end{equation}
	which appears as the argument of the permittivity $\epsilon(\rho)$ in the velocity dispersion profile of Eq.~\eqref{eq:Vrms_Model_RG}.
	
	We estimated the mass profile of the stars as
	\begin{equation}
	\label{eq:M_Stars}
		M_*(r) = \Upsilon L_*(r)\, ,
	\end{equation}
	where $L_*(r)$ is the cumulative luminosity profile, Eq.~\eqref{eq:L_Stars}, and $\Upsilon$ is the stellar mass-to-light ratio that we assumed to be independent of $r$. In our dynamical model, $\Upsilon$ is a free parameter whose prior is described in Sect.~\ref{sec:MCMC}.
	
	The mass profile of the gas, $M_{\rm g}(r)$, derives from the integration of its density profile. This density profile is indicated in Eq.~\eqref{eq:Two_beta}, for NGC 1407 and NGC 5846, and in Eq.~\eqref{eq:rhor_gas_NGC_4486} for NGC 4486. For each galaxy, the mass of the SMBH, $M_\bullet$, is the value reported in Sect.~\ref{sec:M_SMBH}.

	Figure~\ref{fig:Mass_profiles} shows the mass and the mass density profiles of each component. The stars provide the largest contribution to the total mass of each galaxy over most radial range, except in the very centre, where the SMBH dominates, and in the outskirts of NGC 4486 and NGC 5846, where the gas contribution overcomes the star contribution.
	
	\subsection{The MCMC approach}
	\label{sec:MCMC}
	
	For each galaxy, the dynamical model has seven free parameters: four parameters, $\Upsilon$, $\epsilon_0$, $Q$, and $\rho_{\rm c}$, contribute to the gravitational potential well of the galaxy, and three parameters, $\beta_*$, $\beta_{\rm B}$, and $\beta_{\rm R}$, are specific for each tracer. We note that the contributions of the X-ray emitting gas and the SMBH to the gravitational potential have no free parameters. For convenience, hereafter, we use the parameters $\mathcal{B}_*=-\log_{10}(1-\beta_*)$, $\mathcal{B}_{\rm B}=-\log_{10}(1-\beta_{\rm B})$, $\mathcal{B}_{\rm R}=-\log_{10}(1-\beta_{\rm R})$, and $\mathcal{P}_{\rm c} = \log_{10}[\rho_{\rm c}\,(\mathrm{g}\, \mathrm{cm}^{-3})]$. Tangential and radial orbits correspond to $\mathcal{B}_{\rm t}<0$ and $\mathcal{B}_{\rm t}>0$, respectively, where ${\rm t}=(*,{\rm B},{\rm R})$.
	
	We explored this seven-dimensional parameter space with a MCMC algorithm, where we adopted a Metropolis-Hastings acceptance criterion: the random variate  $\vec{x}_{i+1} = (\Upsilon, \epsilon_0, Q, \mathcal{P}_{\rm c}, \mathcal{B}_*, \mathcal{B}_{\rm B}, \mathcal{B}_{\rm R})$, at step $i+1$ of the chain, is obtained from the probability density $G(\vec{x}\vert\vec{x}_i)$, which depends on the random variate $\vec{x}_i$ at the previous step. For $G(\vec{x}\vert\vec{x}_i)$, we adopted a multi-variate Gaussian density distribution with mean value $\vec{x}_i$. As the standard deviations of this Gaussian distribution, we chose 5\% of the ranges of the uniform priors that we adopted for the parameters of the model.
	
	For each single tracer ${\rm t}=(*,{\rm B},{\rm R})$, we adopted the likelihood function
	\begin{equation}
	\label{eq:Likelihood}
	\mathcal{L}_{\rm t}(\vec{x}) = \left[\prod_{i = 1}^{N_{\rm t}}\frac{1}{\sqrt{2\pi\Delta V^2_{\rm rms, data, {\rm t}}(R_i)}}\right]\exp\left[-\frac{\chi^2_{\rm t}(\vec{x})}{2}\right],
	\end{equation}
	where
	\begin{equation}
	\label{eq:chi2_Stars}
	\chi^2_{\rm t}(\vec{x}) = \sum_{i = 1}^{N_{\rm t}}\frac{\left[V_{\rm rms, mod,{\rm t}}(R_i, \vec{x}) - V_{\rm rms, data,{\rm t}}(R_i)\right]^2}{\Delta V^2_{\rm rms, data,{\rm t}}(R_i)}\, ,
	\end{equation}
	$N_{\rm t}$ is the number of data points of the tracer, $V_{\rm rms, data,{\rm t}}(R_i)$ and $\Delta V_{\rm rms, data,{\rm t}}(R_i)$ are the measured root-mean-square velocity dispersions at the projected distances $R_i$ and their uncertainties, and $V_{\rm rms, mod,{\rm t}}(R_i)$ is the expected root-mean-square velocity dispersion derived from the integral in Eq.~\eqref{eq:Vrms_Model_RG} at the same radius. We estimated this integral by adopting two independent linear grids in the $R$ and the $r$ coordinates. Both grids cover the range  $[10^{-4},4800]$~arcsec, with a step of 4~arcsec. This grid size guarantees that our model includes more than $99.4\%$ of the total stellar luminosity of each galaxy.
	
	To consider the three tracers at the same time, we adopted the joint likelihood function
	\begin{equation}
	\label{eq:Likelihood_tot}
		\mathcal{L}(\vec{x}) = \prod_{{\rm t} = 1}^{3} \mathcal{L}_{\rm t}(\vec{x}) = \frac{1}{\mathcal{N}}\exp\left[-\frac{\chi^2_{\rm tot}(\vec{x})}{2}\right], 
	\end{equation}
	with $\mathcal{N}$ the resulting normalisation factor, and
	\begin{equation}
	\label{eq:chi2tot}
	\chi^2_{\rm tot}(\vec{x}) = \chi^2_*(\vec{x}) + \chi^2_{\rm B}(\vec{x}) + \chi^2_{\rm R}(\vec{x}). 
	\end{equation}
	
	The Metropolis--Hastings ratio is 
	\begin{equation}
	\label{eq:R_MH}
	A = \frac{p(\vec{x}) \times \mathcal{L}(\vec{x})}{p(\vec{x}_i) \times \mathcal{L}(\vec{x}_i)} \frac{G(\vec{x} \vert \vec{x}_i)}{G(\vec{x}_i \vert \vec{x})}\, ,
	\end{equation}
	where $p(\vec{x})$ is the product of the priors of the components of $\vec{x}$. In the case of uniform priors, as we adopted in our analysis, $p(\vec{x}) = $~const. If $A \geq 1$, we accept the proposed combination of free parameters and thus we set $\vec{x}_{i + 1} = \vec{x}$; otherwise, we either set $\vec{x}_{i + 1} = \vec{x}$, with probability $A$, or $\vec{x}_{i + 1} = \vec{x}_i$, with probability $1 - A$. For each galaxy, we ran two Markov chains with two different random seeds. Each chain has $8 \times 10^4$ steps, where the first $6 \times 10^3$ steps are discarded as the burn-in chain. We assessed the convergence of the chains with the~\citet{Gelman_and_Rubin_1992} diagnostics, and the convergence of each chain was further confirmed according to the~\citet{Geweke_1992} diagnostics. In the following, we consider the posterior distributions of the parameters from the two chains combined.

	
	To model each galaxy with our MCMC analysis, we need to adopt a prior for each of our seven free parameters. For NGC 1407, the photometric information available in the $B$-band suggests the uniform prior $[4.0,11.2]$~$M_\odot/L_\odot$ for $\Upsilon$. For NGC 4486 and NGC 5846, with photometric information in the $r$-band, we adopted the uniform prior $[1.7,5.5]$~$M_\odot/L_\odot$. We chose both prior intervals according to the ranges expected from the SPS models of~\citet{Humphrey_2006} and~\citet{Zhang_2007}, for the $B$-band, and of~\citet{Bell_2003} and~\citet{Zibetti_2009}, for the $r$-band. These ranges depend on the initial mass function: for the $B$-band, the lower and upper limits are set by the~\citet{Kroupa_2001} and~\citet{Salpeter_1955} initial mass functions, respectively, and for the $r$-band, the lower and upper limits are set by the~\citet{Bottema_1993} and~\citet{Salpeter_1955} initial mass functions, respectively. For all the three orbital anisotropy parameters, $\mathcal{B}_*$, $\mathcal{B}_{\rm B}$, and $\mathcal{B}_{\rm R}$, we adopted the uniform prior $[-1.5,1.0]$, thus including almost all possible orbits, from very tangential to very radial. For NGC 1407, our priors coincide with those adopted by~\citet{Pota_2015}, who modelled the kinematics of this galaxy in Newtonian gravity with a generalised Navarro-Frenk-White dark matter halo~\citep{Hernquist_1990,Zhao_1996}.
	
	For the three permittivity parameters, we adopted uniform priors in the ranges $(0,1]$, $[0.01,2]$, and $[-27,-23]$, for $\epsilon_0$, $Q$, and $\mathcal{P}_{\rm c}$, respectively. These priors are those adopted in~\citet{Cesare_2020}, except for $\epsilon_0$: our range here is wider, compared to the range $[0.1,1]$ of~\citet{Cesare_2020}. Our range for $\mathcal{P}_{\rm c}$ includes the extreme values adopted by \citet{Matsakos_and_Diaferio_2016}. This range includes the typical galaxy densities where Newtonian gravity does not suffice to describe the kinematic data. 
		In passing, the range $[10^{-27},10^{-23}]$~g~cm$^{-3}$ includes the lowest density of the interstellar medium ($\sim$$10^{-25}$~g~cm$^{-3}$,~\citealt{Wood_and_Linsky_1997,Thornton_1998}), but it is well above the average density of the intergalactic medium  ($\sim$$10^{-31}$~g~cm$^{-3}$,~\citealt{Meiksin_2009}). The smallest pressure reached in ultra-high vacuum chambers on Earth is $\sim$$10^{-10}$~mbar, corresponding to a density $\sim$$10^{-17}-10^{-16}$~g~cm$^{-3}$ for standard air with molar mass 29 g~mol$^{-1}$ \citep{Cooper_2019,Liu_2021}. 
	The priors on the free parameters of the model are summarised in Table~\ref{tab:Vrms_models_parameters}.

	\section{Results}
	\label{sec:Results}
	
    We now illustrate how our model describes the velocity dispersion profiles of the three galaxies (Sect.~\ref{sec:Results_Vrms}) and discuss our estimates of the mass-to-light ratios and the anisotropy parameters (Sect.~\ref{sec:MonL_beta_parameters}) and of the parameters of the RG permittivity (Sect.~\ref{sec:RGParameters}).

	\subsection{Velocity dispersion profiles}
	\label{sec:Results_Vrms} 
	
\begin{figure*}
    	\centering
    	\includegraphics[width=17cm]{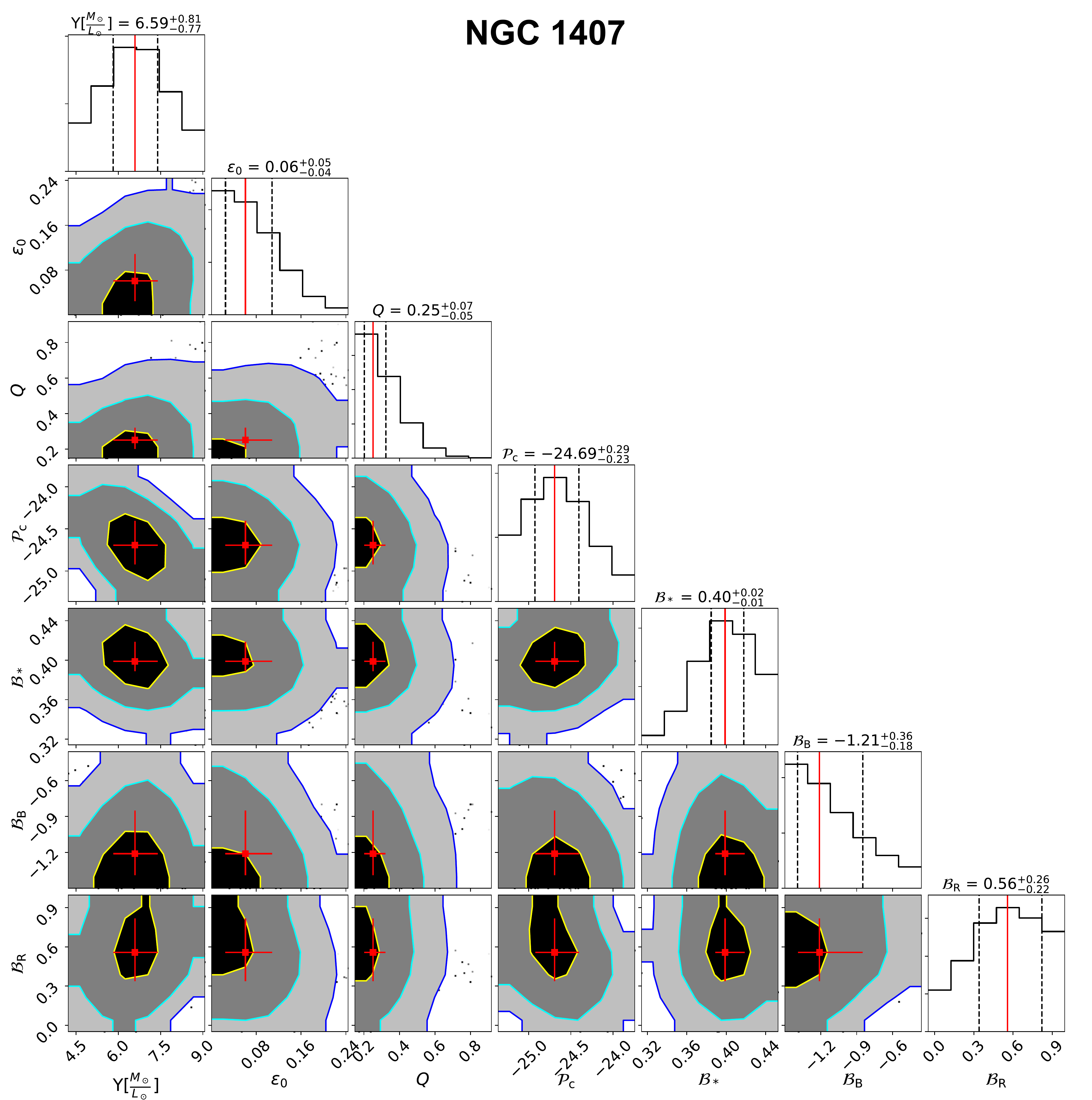}
    	\caption{Posterior distributions of the seven parameters derived from the two joint Markov chains for NGC 1407. The parameters are the stellar mass-to-light ratio, $\Upsilon$, the velocity anisotropy parameters of the three tracers, $\mathcal{B}_*$, $\mathcal{B}_{\rm B}$, $\mathcal{B}_{\rm R}$, and the three permittivity parameters, $\epsilon_0$, $Q$, and $\mathcal{P}_{\rm c}$. The red squares with error bars show the medians and their uncertainites, set by the $15.9$ and $84.1$ percentiles of the posterior distributions. The yellow, cyan, and blue contours limit the 1$\sigma$, 2$\sigma$, and 3$\sigma$ regions, respectively. The medians and their uncertainties are also reported above each column and in the one-dimensional posterior distributions in the \textit{top panels} of each column as red solid and black dashed lines, respectively.}
    	\label{fig:CP_NGC_1407_Final}
    \end{figure*}

\begin{figure*}
    	\centering
    	\includegraphics[width=17cm]{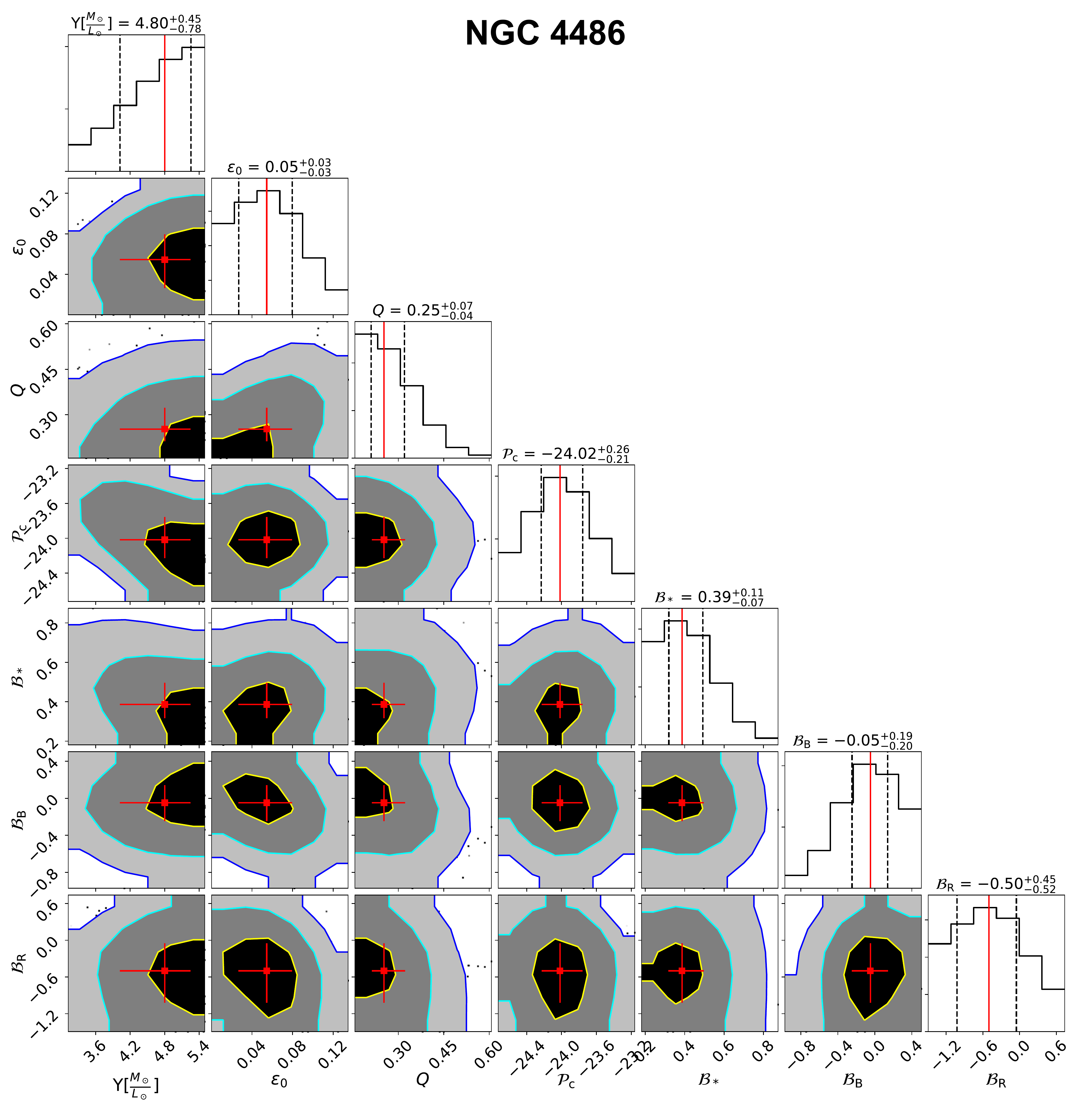}
    	\caption{Same as Fig.~\ref{fig:CP_NGC_1407_Final}, but for NGC 4486.}
    	\label{fig:CP_NGC_4486_Final}
    \end{figure*}

\begin{figure*}
    	\centering
    	\includegraphics[width=17cm]{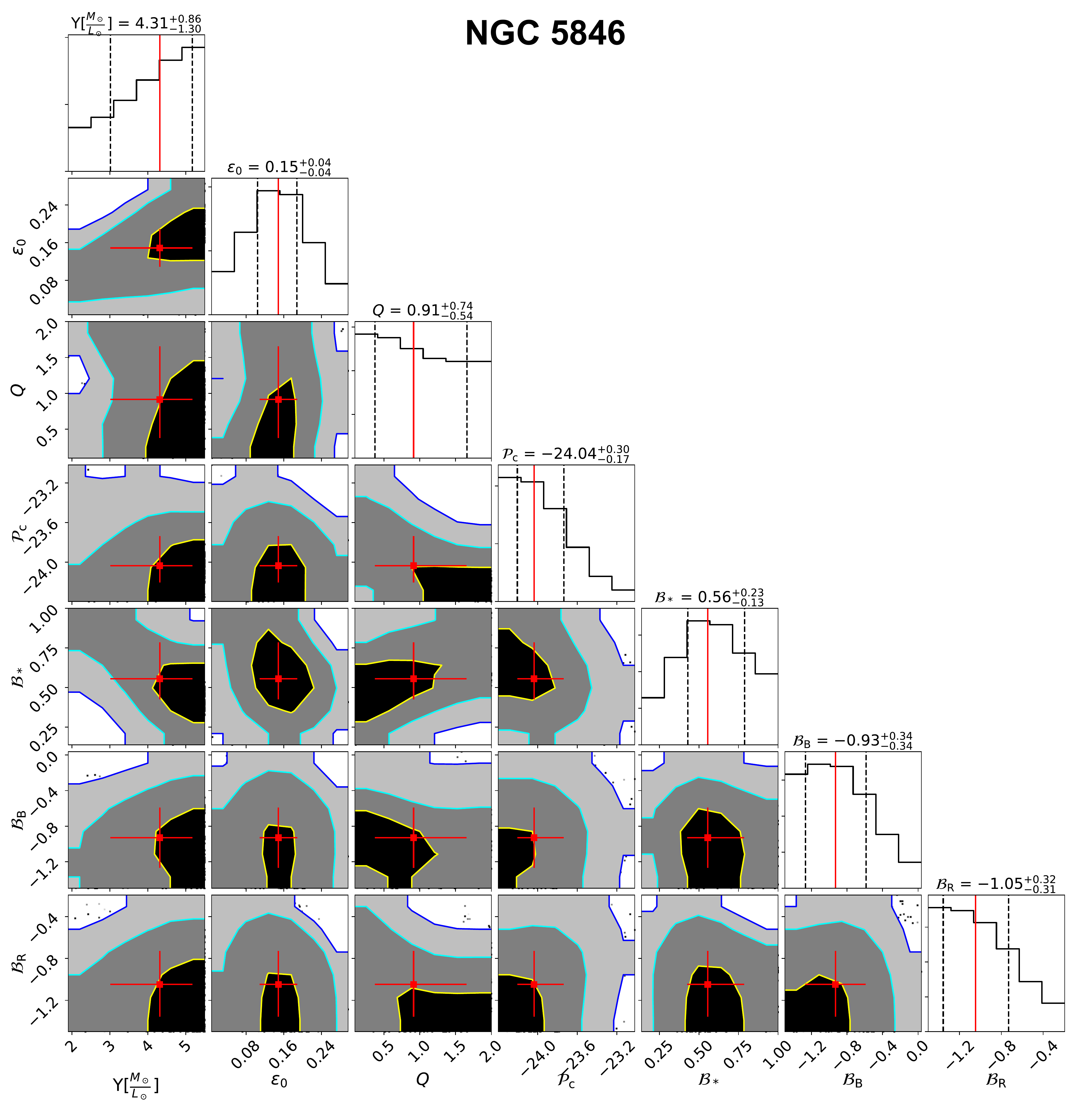}
    	\caption{Same as Fig.~\ref{fig:CP_NGC_1407_Final}, but for NGC 5846.}
    	\label{fig:CP_NGC_5846_Final}
    \end{figure*}

\begin{figure*}
    	\centering
    	\includegraphics[width=17cm]{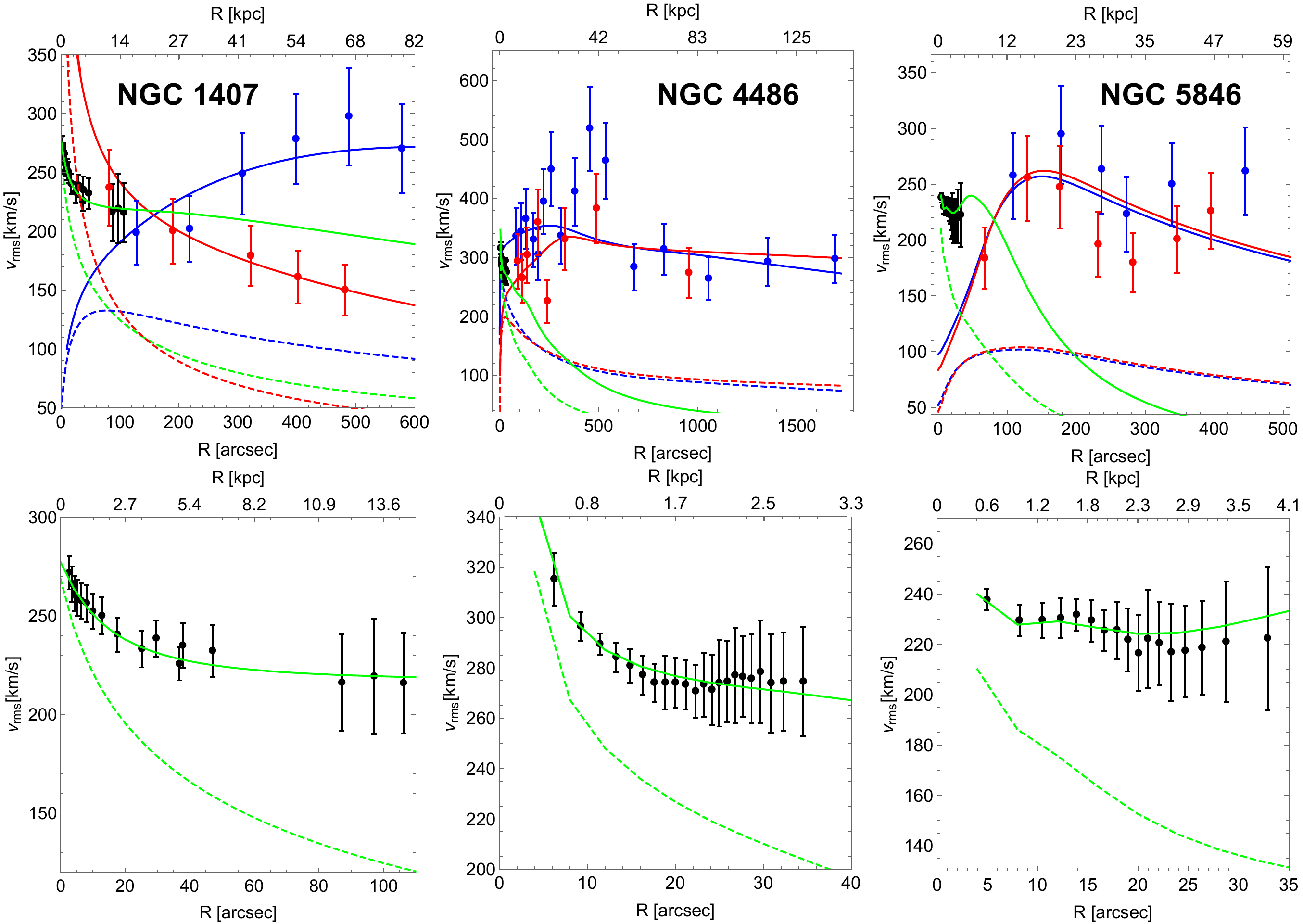}
    	\caption{Models of the root-mean-square velocity dispersion profiles emerging from our MCMC analysis in RG (solid lines). \textit{Upper panels}: the green, blue, and red solid lines show the model profiles for the stars, blue GCs, and red GCs, respectively. \textit{Lower panels}: zoom-in of the stellar profiles. The solid circles with error bars are the measured profiles from Fig.~\ref{fig:vrms_Data_NGC_1407_NGC_4486_NGC_5846}. The green, blue, and red dashed lines show the kinematic profiles of the stars, blue GCs, and red GCs computed with Eq.~\eqref{eq:Vrms_Model} where the values of the mass-to-light ratio $\Upsilon$ and the velocity anisotropy parameters $\beta$ are those of the RG model but the gravitational field is Newtonian with no dark matter. The discrepancy with the RG curves highlights the gravitational boosting generated by the RG gravitational permittivity. }
    	\label{fig:Vrms_models}
    \end{figure*}

    Figures~\ref{fig:CP_NGC_1407_Final}--\ref{fig:CP_NGC_5846_Final} show the posterior distributions of the parameters of the model obtained from the two joint Markov chains for each galaxy. Figure~\ref{fig:Vrms_models} compares the kinematic data with the models of the $V_{\rm rms}$ profiles whose parameters are the medians of the posterior distributions (solid lines). These parameters are listed in Table~\ref{tab:Vrms_models_parameters}. The shape with a single peak of most of the posterior distributions justifies the use of the medians as estimators of the free parameters. 
    
    Figure~\ref{fig:Vrms_models} shows that, globally, the RG models provide an adequate description of the kinematic data of the three tracers for all the three galaxies. The reduced $\chi^2$'s listed in Table~\ref{tab:Vrms_models_parameters} support this conclusion. Nevertheless, in NGC 5846, the model $V_{\rm rms}$ of the blue GCs is systematically smaller than the data; similarly, in NGC 4486, the model is unable to describe the peak, at $R\sim 40$~kpc, of the $V_{\rm rms}$ profile of the blue GCs. This poor modelling might originate from the assumption that these two galaxies are isolated rather than embedded in a larger system, as we discuss below. These discrepancies might also be reduced if we remove our assumption that the velocity anisotropy parameters $\beta$ are independent of radius $r$ or that the galaxies have null net rotation: albeit weak, the net galaxy rotation is not completely absent in the real galaxies. 
    
    \subsection{Mass-to-light ratios and anisotropy parameters}
    \label{sec:MonL_beta_parameters}

    We set the ranges of the uniform prior for the mass-to-light ratios according to the SPS models. For NGC 1407, the posterior distribution of the mass-to-light ratio peaks around the centre of this range. For NGC 4486 and NGC 5846, the posterior distribution of the mass-to-light ratio peaks close to the upper limit of the range of the prior, but the probability that the actual mass-to-light ratio lies in the rest of the range is substantial. Thus, we conclude that RG can indeed describe the data with stellar mass-to-light ratios consistent with the SPS models.
    
    For NGC 1407, the values of $\Upsilon$, $\mathcal{B_*}$, $\mathcal{B_{\rm B}}$, and $\mathcal{B_{\rm R}}$ that we obtained are at $-0.38$$\sigma$, $+1.00$$\sigma$, $-0.21$$\sigma$, and $+0.097$$\sigma$, respectively, from the parameters found by~\citet{Pota_2015} in their analysis with Newtonian gravity and a dark matter halo.
	The RG models require radial orbits for the stars and tangential orbits for the blue GCs in all the three galaxies. The red GCs have tangential orbits in NGC 4486 and NGC 5846. In NGC 1407, the red GCs have radial orbits, in agreement with the result found by \citet{Pota_2015} in Newtonian gravity.

    \subsection{Parameters of the RG permittivity}
	\label{sec:RGParameters}
	
\begin{figure*}
    	\centering
    	\includegraphics[width=17cm]{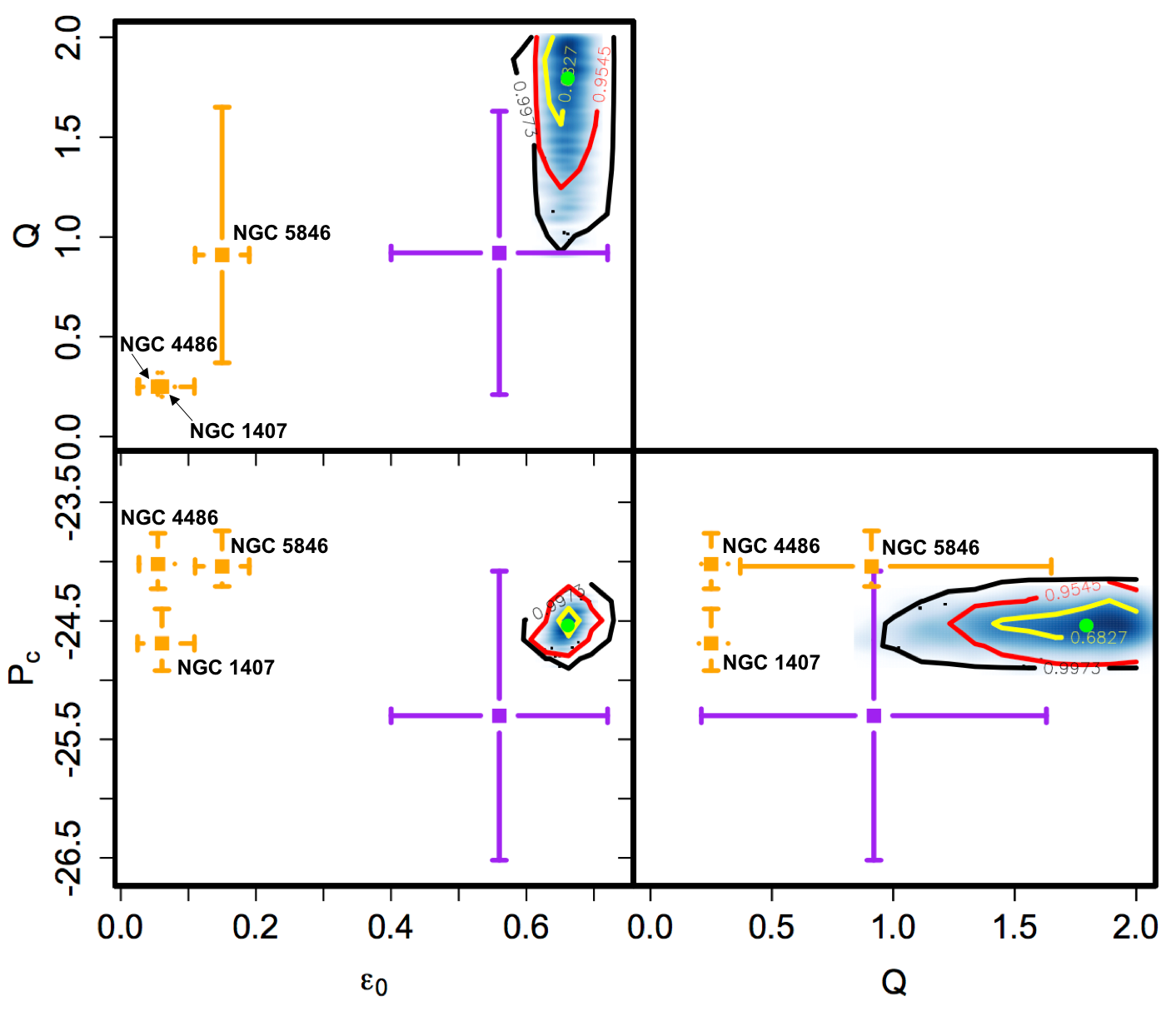}
    	\caption{Permittivity parameters estimated from the three E0 galaxies in our sample (orange squares with error bars) compared with the permittivity parameters estimated by~\citet{Cesare_2020} from the DMS disk galaxies. The purple squares with error bars show the means of the permittivity parameters found for the individual DMS galaxies. The light blue shaded areas show the posterior distributions of the three permittivity parameters found with an approximate procedure from the entire DMS sample, with the green dots indicating their median values and the yellow, red, and black contours indicating the 1$\sigma$, 2$\sigma$, and 3$\sigma$ levels, respectively.}
    	\label{fig:Scatter_plots_RG}
    \end{figure*}

    The orange squares with error bars in Fig.~\ref{fig:Scatter_plots_RG} show the permittivity parameters estimated for the three E0 galaxies: the parameters are consistent with each other within 2$\sigma$, suggesting their universality. Their mean values are $\{\epsilon_0,Q,\mathcal{P}_{\rm c}\} = \{0.089^{+0.038}_{-0.035}, 0.47^{+0.29}_{-0.21}, -24.25^{+0.28}_{-0.20}\}$. Figure~\ref{fig:Scatter_plots_RG} also shows the permittivity parameters derived from 30 disk galaxies of the DMS sample~\citep{Cesare_2020}: the purple squares with error bars show the means of the parameters estimated for each individual DMS galaxy, whereas the green dots at the centre of the shaded areas show the global values derived with an approximate procedure from the entire sample of 30 galaxies.
 
    Our estimates of $\mathcal{P}_{\rm c}$ and $Q$ are within 1$\sigma$ from the mean values derived from the individual DMS galaxies, whereas the mean DMS permittivity of the vacuum $\epsilon_0$ is within 3$\sigma$ from our three estimates. This marginal discrepancy might originate from our simplistic modelling of the E0 galaxies. Indeed, here we assume that these galaxies do not have net rotation and are relaxed and isolated, whereas, in fact, these galaxies are in clusters or groups and show signs of interactions with nearby galaxies: NGC 1407 is at the centre of the Eridanus A group~\citep{Brough_2006} and certainly feels the gravitational  field of its neighbouring galaxy NGC 1400. 
    NGC 4486, namely M87, is the central giant elliptical galaxy of the Virgo cluster~\citep{Strader_2011} and NGC 5846 is the central and brightest galaxy of a galaxy group~\citep{Mahdavi_2005}. Moreover, the GC sample of NGC 5846 might be contaminated by the GCs of its neighbour NGC 5846A~\citep{Pota_2013}. The environment is expected to have a relevant effect on the intensity of the gravitational field in RG~\citep{Matsakos_and_Diaferio_2016}; this effect clearly propagates into the actual value of the permittivity of the vacuum $\epsilon_0$. Environmental effects can also be responsible for the poor modelling of the velocity dispersion profiles of the blue GCs of NGC 4486 and NGC 5846 shown in Fig.~\ref{fig:Vrms_models}. 
    
    Assessing the gravitational interactions of neighbouring galaxies in RG is difficult and is beyond the purpose of our present work. The shape of the mass density distribution of an individual source determines the refraction angle of the lines of the gravitational field and thus how the direction and the intensity of the gravitational field depend on the vector $\vec{r}$ of the distance from the source \citep{Matsakos_and_Diaferio_2016}. Therefore, estimating the gravitational interactions among galaxies with arbitrary mass density distributions requires an accurate integration of the field equations with little room for simplyfing assumptions.
 
    The tension between our vacuum permittivity $\epsilon_0$ and the vacuum permittivity derived from the DMS sample might become more severe if we consider the DMS global value, namely the green dot in Fig.~\ref{fig:Scatter_plots_RG}. Our $\epsilon_0$ shows a $14.8\sigma$ tension from the DMS parameter, which is clearly driven by the small width of the posterior distribution of the DMS global value. On the contrary, the parameter $Q$ is within $3.4\sigma$ from the DMS global parameter. More importantly, the critical density $\mathcal{P}_{\rm c}$, which represents the most relevant parameter of RG, is within $1.4\sigma$ from the DMS global parameter. The  large discrepancy for $\epsilon_0$ might be due to the approximate procedure adopted by \citet{Cesare_2020} to estimate this DMS global value: with their MCMC analysis, \citet{Cesare_2020} only explored the space of the three permittivity parameters while keeping fixed the mass-to-light ratio $\Upsilon$ and the disk-scale height $h_z$ of each of the 30 individual galaxies. This approach is a shortcut to the appropriate, but computationally overwhelming, procedure of exploring the 63-dimensional parameter space of the full sample, namely the two parameters, $\Upsilon$ and $h_z$, of each of the 30 galaxies and the three permittivity parameters. The set of the permittivity global parameters returned by this shortcut still describes the kinematic profiles of each galaxy, but the agreement is poorer than the agreement obtained by modelling each galaxy individually. Therefore, these values of the permittivity parameters should be considered with caution. 
 
    \begin{table*}
    	\caption{Priors and parameters of the $V_{\rm rms}(R)$ model, Eq.~\eqref{eq:Vrms_Model_RG}, estimated from the three dynamical tracers.}
    	\label{tab:Vrms_models_parameters}	
    	\centering
    	\begin{tabular}{lcccccccc}
    		\hline\hline
    		NGC & $\Upsilon$& $\epsilon_0$ & $Q$ & $\mathcal{P}_{\rm c}$ & $\mathcal{B}_*$ & $\mathcal{B}_{\rm B}$ & $\mathcal{B}_{\rm R}$ & $\chi^2_{\rm red, tot}$\\
    		& $\left[\frac{M_\odot}{L_\odot}\right]$ & & &  &&&&  \\
    		(1) & (2) & (3) & (4) & (5) & (6) & (7) & (8) & (9)\\
    		\hline
    		& & $(0.0,1.0]$ & $[0.010,2.0]$ & $[-27,-23]$ & $[-1.5,1.0]$ & $[-1.5,1.0]$ & $[-1.5,1.0]$ & \\
    		1407 & $6.59^{+0.81}_{-0.77}$ & $0.061^{+0.048}_{-0.036}$ & $0.25^{+0.07}_{-0.05}$ & $-24.69^{+0.29}_{-0.23}$ & $0.40^{+0.02}_{-0.01}$ & $-1.21^{+0.36}_{-0.18}$ & $0.56^{+0.26}_{-0.22}$ & $0.20$\\
    		4486 & $4.80^{+0.45}_{-0.78}$ & $0.055^{+0.025}_{-0.028}$ & $0.25^{+0.07}_{-0.04}$ & $-24.02^{+0.26}_{-0.21}$ & $0.39^{+0.11}_{-0.07}$ & $-0.048^{+0.19}_{-0.20}$ & $-0.50^{+0.45}_{-0.52}$ & $0.79$\\
    		5846 & $4.31^{+0.86}_{-1.30}$ & $0.15^{+0.04}_{-0.04}$ & $0.91^{+0.74}_{-0.54}$ & $-24.04^{+0.30}_{-0.17}$ & $0.56^{+0.23}_{-0.13}$ & $-0.93^{+0.34}_{-0.34}$ & $-1.05^{+0.32}_{-0.31}$ & $0.63$\\
    		\hline
    	\end{tabular}
    	\tablefoot{Column 1: galaxy name; Cols. 2--8: parameters of the model; Col. 9: reduced chi-square, $\chi_{\rm red, tot}^2$. The first row lists the ranges of the uniform priors on the parameters of the model and the remaining rows list the estimated parameters with their uncertanties. The uniform prior on the mass-to-light ratio has range $[4.0,11.2]$~$M_\odot/L_\odot$ for NGC 1407, and  $[1.7,5.5]$ for NGC 4486 and NGC 5846.}
    \end{table*}
    
	\section{Discussion and conclusion}
	\label{sec:Conclusions}
	
	We used the kinematic information of the stellar population and of the GCs in the outer regions of three E0 galaxies from the SLUGGS survey~\citep{Pota_2013,Brodie_2014,Forbes_2017} to show that RG can describe the kinematics of spherical systems without requiring the existence of dark matter. Our results complement the ability of RG to describe the rotation curves and the vertical velocity dispersion profiles of 30 disk galaxies from the DMS~\citep{DMSI_2010a} shown in \citet{Cesare_2020}. In RG, dark matter is mimicked by the gravitational permittivity, a monotonic function of the local mass density that is expected to be universal. For the permittivity, according to \citet{Matsakos_and_Diaferio_2016} and \citet{Cesare_2020}, we adopted a smooth function that depends on three parameters: the permittivity of the vacuum $\epsilon_0$, the critical density $\rho_{\rm c}$, and the transition slope $Q$. 
	
	We find that the sets of parameters of the three E0 galaxies are consistent with each other within 2$\sigma$. Their values averaged over the three galaxies are  $\{\epsilon_0,Q,\mathcal{P}_{\rm c}\} = \{0.089^{+0.038}_{-0.035}, 0.47^{+0.29}_{-0.21}, -24.25^{+0.28}_{-0.20}\}$, with  $\mathcal{P}_{\rm c} = \log_{10}[\rho_{\rm c}\,(\mathrm{g}\, \mathrm{cm}^{-3})]$. With this permittivity, RG requires stellar mass-to-light ratios in agreement with SPS models and tangential or radial orbits for the GCs, depending on the galaxy and the colour of the GCs. In particular, for NGC 1407, the mass-to-light ratio and the orbital anisotropy parameters are within 1$\sigma$ from the values found by~\citet{Pota_2015}, who modelled this galaxy with Newtonian gravity and a dark matter halo. 
	
	By modelling the kinematic properties of each individual DMS disk galaxy, \citet{Cesare_2020} derived the mean values of the permittivity parameters $\{\epsilon_0,Q,\mathcal{P}_{\rm c}\} = \{0.56 \pm 0.16, 0.92 \pm 0.71, -25.30 \pm 1.22\}$. The mean values of $Q$ and $\mathcal{P}_{\rm c}$ are within 1$\sigma$ from the parameters we find here, whereas $\epsilon_0$ is within 3$\sigma$. These combined results support the universality of the parameters of the RG permittivity.
	
   \citet{Cesare_2020} also derived a set of global parameters $\{\epsilon_0,Q,\mathcal{P}_{\rm c}\} = \{0.661^{+0.007}_{-0.007}, 1.79^{+0.14}_{-0.26}, -24.54^{+0.08}_{-0.07}\}$ with a MCMC analysis where the mass-to-light ratio and the disk thickness of each individual galaxy were kept fixed. This simplified approach returns an approximate estimate of the unique set of the permittivity parameters capable of describing the entire sample of the 30 DMS disk galaxies. This analysis of \citet{Cesare_2020} is suggestive of the possible universality of the permittivity parameters, but their actual values should be taken with caution, because of the approximate procedure used to derive them. Nevertheless, the global DMS values of $\mathcal{P}_{\rm c}$ and $Q$ found by~\citet{Cesare_2020} are within $1.4\sigma$ and $3.4\sigma$, respectively, from the values we find here and thus are still consistent with the expected universality.
   This consistency is particularly important for the critical density $\mathcal{P}_{\rm c}$, which sets the density scale of the transition between the Newtonian and the RG regimes. On the contrary, the global DMS $\epsilon_0$ is 14.8$\sigma$ discrepant from the mean value that we find here. This discrepancy is mostly driven by the width of the posterior distribution found by \citet{Cesare_2020} which is small when compared to the uncertainties of the corresponding mean values (Fig.~\ref{fig:Scatter_plots_RG}). If this width is not severely underestimated, this discrepancy might highlight at least three different possible problems: a too simplistic model of the ellipticals, an incorrect form of the permittivity, or a fundamental fault of RG.
   
   The model of the ellipticals that we adopt here is oversimplified, because we treat the galaxies as isolated systems and completely neglect the net rotation of the galaxies. Indeed, NGC 1407 and NGC 5846 are within galaxy groups and NGC 4486 is the central galaxy of the Virgo cluster. Neglecting the galaxy environment can clearly overlook relevant interactions with neighbouring galaxies  that could actually affect the parameters of the permittivity. Including these environmental effects requires modelling all the galaxies within the system at the same time, with each galaxy modelled according to its morphological type and shape. Properly modelling such a system is arduous and its non-linear effects are difficult to assess at the present stage. Thus, we cannot exclude that environmental effects might be part, if not all, of the cause of the possible tension on $\epsilon_0$ that we find here. Alternatively, the explicit form of the permittivity $\epsilon(\rho)$ (Eq.~\eqref{eq:eps}), which is chosen arbitrarily, might actually be inappropriate, especially in the very low-density regime: hints from the ultra-weak field limit of the covariant version of RG~\citep{Sanna_2021} might suggest a better motivated permittivity that might alleviate this tension. Finally, all of the above might turn out to play a negligible role, and this possible tension on the permittivity of the vacuum might simply betray a fundamental fault of RG.

     \begin{acknowledgements}
     	We sincerely thank the referee who identified a few errors in our original analysis and suggested relevant improvements. We also thank Xavier Hern{\'a}ndez, Federico Lelli, Filippo Fraternali, and Heng Yu for their precious advice and detailed comments that improved the presentation of this work. We acknowledge the help of Alessandro Paggi and Valentina Missaglia in modelling the gas of NGC 5846 and the number density profiles of the GCs. We also thank Michele Cappellari, for his help in the derivation of the stellar root-mean-square velocity dispersion profiles of NGC 4486 and NGC 5846. We thank Compagnia di San Paolo (CSP) for funding the graduate-student fellowship of VC. We also acknowledge partial support from the INFN grant InDark and the Italian Ministry of Education, University and Research (MIUR) under the Departments of Excellence grant L.232/2016. This research has made use of NASA’s Astrophysics Data System Bibliographic Services.
     \end{acknowledgements}
     
     \bibliographystyle{aa}
     \bibliography{bib_paper3_RR}

     \begin{appendix}
     	
     \section{Parameters of the stellar surface brightness profiles of NGC 4486 and NGC 5846}
     \label{sec:SB_stars_NGC_4486_NGC_5846_A}
     
     \begin{table*}
     	\caption{Parameters of the stellar surface brightness profile of NGC 4486.}
     	\label{tab:SLUGGS_Phot_Stars_NGC_4486}	
     	\centering
     	\begin{tabular}{cc}
     		\hline\hline
     		$L_k$  & $\sigma_k$ \\
     		$[L_{\odot}]$ & [arcsec] \\
     		(1) & (2) \\
     		\hline
     		$3.26333\times 10^7$ & 0.31847133 \\
     		$6.31462\times 10^7$ & 0.85316859 \\
     		$1.42502\times 10^8$ & 1.9878101 \\
     		$1.20241\times 10^9$ & 4.3688105 \\
     		$2.92993\times 10^9$ & 7.0315581 \\
     		$6.16303\times 10^9$ & 11.941329 \\
     		$1.08416\times 10^{10}$ & 21.809295 \\
     		$2.11184\times 10^{10}$ & 48.869046 \\
     		$3.15170\times 10^{10}$ & 120.71397 \\
     		\hline
     	\end{tabular}
     \end{table*}
     
     \begin{table*}
     	\caption{Parameters of the stellar surface brightness profile of NGC 5846.}
     	\label{tab:SLUGGS_Phot_Stars_NGC_5846}	
     	\centering
     	\begin{tabular}{cc}
     		\hline\hline
     		$L_k$  & $\sigma_k$ \\
     		$[L_{\odot}]$ & [arcsec] \\
     		(1) & (2) \\
     		\hline
     		$3.09594\times 10^8$ & $1.3004914$ \\
     		$6.31938\times 10^8$ & $2.0720509$ \\
     		$2.52404\times 10^9$ & $4.7753466$ \\
     		$2.41652\times 10^9$ & $8.9026766$ \\
     		$8.05893\times 10^9$ & $16.353135$ \\
     		$9.26616\times 10^9$ & $32.666416$ \\
     		$2.29889\times 10^{10}$ & $78.813294$ \\
     		\hline
     	\end{tabular}
     \end{table*}
     
     Tables~\ref{tab:SLUGGS_Phot_Stars_NGC_4486} and~\ref{tab:SLUGGS_Phot_Stars_NGC_5846} list the parameters of the model of the stellar surface brightness that we adopted for the galaxies NGC 4486 and NGC 5846. These parameters were determined by~\citet{ATLAS_XXI} by fitting the measured surface brightness with the MGE model (Sect.~\ref{sec:SB_stars_NGC_4486_NGC_5846}).

     \section{Uncertainty on the root-mean-square velocity dispersions of  GCs}
     	\label{sec:Errors_Vrms_data_GCs}
     	
     	We adopted the procedure of~\citet{Danese_1980} to estimate the uncertainties $\Delta V_{\rm rms}(R)$ on the values of the root-mean-square velocity dispersion profile $V_{\rm rms}(R)$ that we computed with Eq.~\eqref{eq:Vrms_GCs}, where we considered the GCs in different radial bins. In the reference frame of the centre of mass of the galaxy, the line-of-sight velocity of the $i$-th GC is 
     	\begin{equation}
     	\label{eq:v_Parallel_Danese_1980_Eq5}
     		v_{||,i} = \frac{V_{{\rm rad},i} - V_{\rm sys}}{1 + {V_{\rm sys}}/{c}},
     	\end{equation}
     	where $V_{{\rm rad},i}$ is the observed radial velocity of the GC, $V_{\rm sys}$ is the systemic velocity of the galaxy, and $c = 3 \times 10^5$~km~s$^{-1}$ is the speed of light. The three E0 galaxies of our analysis have redshift $z<0.007$ and the denominator of Eq.~\eqref{eq:v_Parallel_Danese_1980_Eq5} can be approximated to 1. The variance of the line-of-sight velocities of the GCs in each radial bin of radius $R$ is
     	\begin{equation}
     	\label{eq:sigma_Parallel_Danese_1980_Eq6}
     		\sigma_{||}^2(R) = \frac{1}{N-1}\sum_{i = 1}^N v_{||,i}^2\, ,
     	\end{equation}
     	where $N$ is the number of GCs in the bin. This equation neglects the errors on $V_{\rm rad}$ and $V_{\rm sys}$.
     	
     	The quantity
     	\begin{equation}
     	\label{eq:S_Danese_1980}
     		S^2(R) = (N - 1)\frac{\sigma_{||}^2(R)}{V_{\rm rms}^2(R)}
     	\end{equation}
     	is a $\chi_\nu^2$ random variate with $\nu = N - 1$ degrees of freedom. We define $S_-^2$ and $S_+^2$ with the probability $\alpha$ of having $S^2$ in the range  $(S_-^2,S_+^2)$. The value $\alpha=0.68$ defines the upper, indicated with the $+$ sign, and lower, indicated with the $-$ sign, 1$\sigma$ uncertainties on the dispersion of the line-of-sight velocities derived with Eq.~\eqref{eq:sigma_Parallel_Danese_1980_Eq6}. The values $S_-^2$ and $S_+^2$ provide the following estimate of the uncertainty of $V_{\rm rms}(R)$ appearing in Eq.~\eqref{eq:Vrms_GCs}
     	\begin{equation}
     	\label{eq:Deltasigmaminusplus_Parallel_Danese_1980_Eq10}
     		\Delta V_{{\rm rms}, \pm}^2(R) = \left[\left(\frac{\nu}{S_\mp^2(R)}\right)^{1/2} - 1\right]^2\sigma_{||}^2(R) + \frac{\widebar{\delta^2_*}}{N}\left(1 + \frac{\widebar{\delta^2_*}}{2\sigma_{||}^2(R)}\right),
     	\end{equation}
     	where $\widebar{\delta^2_*}$ is
     	\begin{equation}
     	\label{eq:deltameanstar2_Danese_1980}
     		\widebar{\delta^2_*} = \frac{\Delta V_{\rm sys}^2}{\left(1 + {V_{\rm sys}}/{c}\right)^2}
     	\end{equation}
     	and $\Delta V_{\rm sys}$ is the uncertainty on $V_{\rm sys}$. In our analysis, $V_{\rm sys} \ll c$, and thus we can replace $\widebar{\delta^2_*}$  with $\Delta V_{\rm sys}^2$. 
     	In Sect.~\ref{sec:V_rms_GCs_NGC_4486_NGC_5846}, we used the symmetrised uncertainty
     	\begin{equation}
     		\Delta V_{\rm rms}(R) = \frac{\Delta V_{{\rm rms}, -}(R) + \Delta V_{{\rm rms}, +}(R)}{2}\, .
     	\end{equation}
     	
     \end{appendix}
     	
     \listofobjects
	
\end{document}